\documentclass[oneside,11pt,reqno]{amsart}

\pdfoutput=1
\usepackage{hyperref}
\usepackage[capitalize]{cleveref} 
\usepackage[margin=2cm]{geometry}                		
\usepackage{subfigure}
\usepackage{graphicx}			
\usepackage{amsmath}
\usepackage{amssymb}
\usepackage{color}
\usepackage{bm}
\usepackage{comment}
\usepackage{subcaption}
\usepackage{mathtools}
\usepackage{amsaddr}
\usepackage{xfrac}
\usepackage{xcolor}
\usepackage[normalem]{ulem}
\newcommand{\ket}[1]{\left| #1 \right>} 
\newcommand{\bra}[1]{\left< #1 \right|} 
 
\hypersetup{
    colorlinks=true,
    linkcolor=blue,
    citecolor=blue,
    filecolor=blue,
    urlcolor=blue,
}
\newcommand{\rmi}{\mathrm{i}}
\newcommand{\rmd}{\mathrm{d}}
\newcommand{\tr}{\mathrm{tr}}

\newtheorem{result}{Result}

\theoremstyle{remark}

\usepackage{framed}
\definecolor{shadecolor}{gray}{0.95}
\usepackage{cite}

\usepackage[T1]{fontenc}
\usepackage{tikz-cd}
\usetikzlibrary{arrows}
\usetikzlibrary{shapes.arrows}
\usetikzlibrary{decorations.markings}
\usetikzlibrary{decorations.pathreplacing}
\usetikzlibrary{shapes.geometric}
\usetikzlibrary{backgrounds}
\usetikzlibrary{decorations.pathmorphing}
\tikzset{snake it/.style={decorate, decoration=snake}}
\tikzset{tensor/.style={thick,minimum size=.15cm,rectangle,draw, fill=white}}
\tikzset{ shorten <>/.style={ shorten >=#1, shorten <=#1 } }

\title[Matrix-product operator dualities in integrable lattice models]{Matrix-product operator dualities in \\ integrable lattice models}

\begin{document}
\author{Yuan Miao}
\address{Kavli Institute for the Physics and Mathematics of the Universe (WPI), The University of Tokyo Institutes for Advanced Study, The University of Tokyo, Kashiwa, Chiba 277-8583, Japan}
\author{Andras Molnar}
\address{University of Vienna, Faculty of Mathematics, Oskar-Morgenstern-Platz 1, 1090 Vienna, Austria}
\author{Nick G. Jones}
\address{St John’s College and Mathematical Institute, University of Oxford, UK}
\thanks{The published version of this article is J. Phys. A: Math. Theor. 59 305002 (2026); \href{https://doi.org/10.1088/1751-8121/ae880b}{https://doi.org/10.1088/1751-8121/ae880b}}
\begin{abstract}
Matrix-product operators (MPOs) appear throughout the study of integrable lattice models, notably as the transfer matrices. They can also be used as transformations to construct dualities between such models, both invertible (including unitary) and non-invertible (including discrete gauging). We analyse how the local Yang--Baxter integrable structures are modified under such dualities. We see that the $\check{R}$-matrix, that appears in the baxterization approach to integrability, transforms in a simple manner. We further argue for a broad class of MPO dualities that the usual Yang--Baxter $R$-matrix should be extended. This extended $R$-matrix satisfies modified algebraic relations, including a modified Yang--Baxter algebra previously identified in the unitary case, that altogether give a local integrable structure underlying the commuting transfer matrices of the dual model. We illustrate these results with two case studies, analysing an invertible unitary MPO and a non-invertible MPO both applied to the canonical XXZ spin chain. The former is the cluster entangler, arising in the study of symmetry-protected topological phases, while the latter is the Kramers--Wannier duality. We show several results for MPOs with exact MPO inverses that are of independent interest.
\end{abstract}
\maketitle
\tableofcontents

\newpage

\section{Introduction}
Integrable spin chains are key to our understanding of quantum many-body systems, giving us analytical tools to explore wide-ranging phenomena \cite{Baxter16,Gaudin_2014,Korepin97,Jones03}. For example, consider the spin-1/2 XXZ chain with Hamiltonian
\begin{align}
H_{\mathrm{XXZ}}= \sum_{j=1}^{\ell} \Big( X_j X_{j+1} + Y_j Y_{j+1} +\Delta (\mathbb{I}+Z_j Z_{j+1} ) \Big)     \qquad \Delta \in \mathbb{R} \label{eq:XXZ} \ ;
\end{align}
the local terms are products of the usual Pauli operators on each site, and $H_{\mathrm{XXZ}}$ has a $U(1)$ spin rotation symmetry generated by $\frac{1}{2}\sum_j Z_j$. This model describes the transition between certain symmetry-breaking phases, as well as a gapped paramagnetic phase \cite{Sachdev11}. For $\lvert\Delta\rvert< 1$ it is described at low energies by the compactified free boson conformal field theory (CFT), and tuning $\Delta$ allows us to explore the moduli space of this CFT on the lattice \cite{Alcaraz_PRL_1987, Ginsparg88}. 

\begin{figure}
    \centering
\includegraphics[]{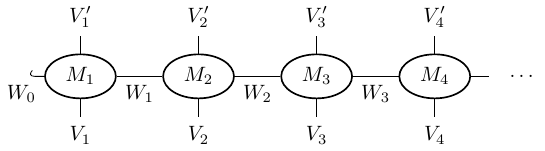}
\caption{A reference matrix-product operator (MPO). Each four-index tensor $M_j$, that may also depend on some parameters, has two physical indices (vertical lines) and two virtual indices (horizontal lines). For fixed virtual indices, $M_i$ gives a linear transformation between physical Hilbert spaces $V_i^{}\rightarrow V_i'$. These are referred to as `quantum space'(s) in some integrability literature. Fixing the physical indices $M_i$ gives a matrix mapping $W_{i-1}\rightarrow W_i$. The $V_i$ are referred to in the literature as bond, virtual or auxilliary spaces. We will often take translation-invariant MPOs acting on a fixed Hilbert space. In this case all $M_i=M$, all virtual spaces are isomorphic, and, taking the trace over the virtual space $V_0\equiv W_j$, we have an operator from $V^{\otimes \ell}$ to itself.}\label{fig:MPO}
\end{figure}

The XXZ spin-chain is solvable using the Bethe ansatz~\cite{Bethe_1931, Hulthen_1938}, and the underlying integrable structure is a one-parameter solution to the Yang--Baxter equation called the $R$-matrix. This solution corresponds to the six-vertex model of classical statistical mechanics \cite{Baxter16, Gaudin_2014}. From the $R$-matrix we can construct commuting matrix-product operators (MPOs, see \cref{fig:MPO}), known as transfer matrices, from which one can derive the Hamiltonian and local conserved charges of $H_{\mathrm{XXZ}}$. 
Note that even with this integrable structure, computing correlation functions remains highly non-trivial \cite{Korepin97, JimboMiwa1995}. 
Since transfer matrices are expressed as MPOs, much of the formalism related to integrable models can be understood in the language of tensor networks \cite{Katsura:2009vx, Murg12}.

In this work we are interested in maps or dualities between different spin chains \cite{Kramers41,Cobanera10,Aasen16,Aasen20,Jones22}. In particular, we seek to understand the interplay between these maps and the underlying local integrable structure of the spin chain, such as the ${R}$-matrix. Many interesting examples of such dualities are known, such as the Kramers--Wannier duality in the Ising model \cite{Kramers41}. This particular duality can be understood more generally as gauging the $\mathbb{Z}_2$ symmetry in a spin chain \cite{Kogut79,Borla21,Seiberg:2023cdc, Seiberg:2024gek} or as a `vertex-face correspondence' in the related classical statistical mechanical model~\cite{Baxter:1972wh, Kojima:2005qz, Ikhlef:2016sgm}.

Importantly, the Kramers--Wannier duality can itself be expressed as an MPO \cite{Aasen16}. We will be interested in the effect of MPO transformations applied to integrable spin chains more generally. The simplest such transformation would be a unitary MPO with bond-dimension one, i.e.  a local on-site unitary transformation. These are trivial and can be easily absorbed into the physical indices of the transfer matrix. It is less clear what happens when we conjugate by a matrix-product unitary (MPU); recall that an MPU is a unitary MPO with bond-dimension greater than one. We will see that these transformations are globally very simple, but change the local integrable structure. Finally, the most interesting case that we consider are MPOs that change the spectrum of the model. These represent
non-invertible duality transformations, generalising the Kramers--Wannier duality.

The MPU case was previously explored in Ref.~\cite{Jones22}, we expand and generalize this discussion here. In particular, we show the extended $R$-matrix defined in that reference satisfies an algebraic relation that is the key local tensor identity underlying the integrability of the dual model. We also show that the MPU analysis applies also to MPOs that have an exact (i.e. finite bond-dimension) MPO inverse. We further explore the connection to Jones' baxterization approach to integrable models \cite{Jones03}. Note that despite their apparent simplicity, certain MPUs are physically important and appear in the setting of symmetry-protected topological (SPT) phases. These MPUs act as SPT entanglers \cite{Tantivasadakarn23,Zhang23}
that map trivial paramagnetic Hamiltonians to SPT Hamiltonians, where the ground states of the latter have non-trivial symmetry fractionalisation \cite{Gu09,Pollmann10,Fidkowski10,Turner11,Schuch11,Chen11,Pollmann12,Senthil13}. Indeed, one motivation to consider these duality transformations is to map between related integrable models with differing symmetry properties and SPT physics, and to identify integrable transitions between SPT phases. 

The outline of the paper is as follows. In \cref{sec:integrable} we review the Yang--Baxter (in terms of quantum group) and baxterization constructions of integrable spin chains and introduce the $R$ and $\check{R}$-matrices that appear in each approach respectively. In \cref{sec:dualities} we introduce three classes of dualities and give a preliminary discussion of their action. We then analyse two cases in detail, first a bond-dimension two MPU in \cref{sec:cluster} and then the non-invertible Kramers--Wannier MPO in \cref{sec:KW}. The first example is representative of general features of MPO with exact MPO inverse. We outline some results for this class of MPO that are of independent interest, and then conclude our discussion of maps and dualities in \cref{sec:conclusion}. 

\section{Review: key notions of integrability and Yang--Baxter equation}\label{sec:integrable}

In this section, we present the key notions of quantum integrability in terms of MPOs. The renowned Yang--Baxter equation (YBE) can be written naturally in this language. After introducing the canonical approach to Yang--Baxter integrability, we show another algebraic approach to quantum integrability called \emph{baxterization}. The latter approach is useful to reveal the integrable structure of models that are related by duality transformations.

\begin{figure}
    \centering
\includegraphics[]{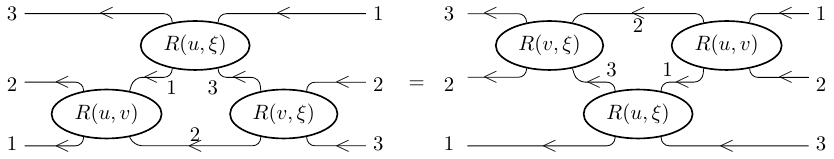}
\caption{`Scattering' YBE for $R$-matrix. The labels $1$, $2$ and $3$ stands for the three vector spaces in $V \otimes V \otimes V$. Arrows indicate the operators acting from right to left.}
    \label{fig:RYBE}
\end{figure}

\subsection{The $R$ and $\check{R}$-matrices}

The $R$-matrix is a four-index tensor $R_{\alpha,\beta}^{\gamma,\delta}(u,v)$ that acts on $V\otimes V$, for some `bond' vector space $V$, by $R(u,v) \ket{\alpha\beta} = \sum_{\gamma\delta} R_{\alpha,\beta}^{\gamma,\delta}(u,v) \ket{\gamma\delta}$, and depends on two spectral parameters $u$ and $v$. Abusing notation, define $R_{12}= R\otimes\mathbb{I}$, $R_{23}= \mathbb{I}\otimes R$, and $R_{13}=\sum_{i} x_i \otimes \mathbb{I} \otimes y_i $ where $R=\sum_i x_i \otimes y_i$; then the YBE is given by \cite{Baxter16, Murg12} 
\begin{align}
R_{12}(u,v) R_{13}(u,\xi)R_{23}(v,\xi) = R_{23}(v,\xi)    R_{13}(u,\xi)  R_{12}(u,v)  \; .
\label{eq:YBE_123}
\end{align}
Diagrammatically this is given in \cref{fig:RYBE}, and we call it the scattering YBE.

\begin{figure}
    \centering
\includegraphics[]{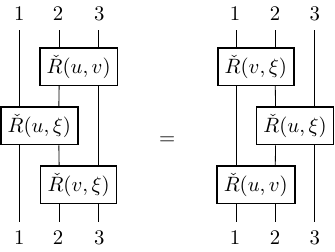}
\caption{`Circuit' YBE for $\check{{R}}$-matrix. This is \cref{eq:checkYBE} where we specialize to the case where $\check{{R}}$ acts on two neighbouring sites.  This moreover represents \cref{eq:check_R_YBE} when we take the representation over a tensorized Hilbert space where $\check{{\mathcal{R}}}_j \mapsto \check{R}_{j,j+1}$.} 
    \label{fig:RcheckYBE}
\end{figure}

Alternatively, we define $\check{R}(u,v) = {SW} \, R(u,v)$, where ${SW}$ is the swap operator (also known as the permutation operator) ${SW}: V_1 \otimes V_2 \to V_2 \otimes V_1$. The operators $\check{R}$ satisfy the equation
\begin{align}
\check{R}_{23}(u,v) \check{R}_{12}(u,\xi)\check{R}_{23}(v,\xi) = \check{R}_{12}(v,\xi) \check{R}_{23}(u,\xi)    \check{R}_{12}(u,v)  \ .              \label{eq:checkYBE}
\end{align}
Diagrammatically this is given in \cref{fig:RcheckYBE}, and we call it the circuit YBE. In the following we allow $\check{R}$ to act on $k>2$ sites, in which case we write $\check{R}_i$ for the operator that acts on sites $[i, i+1, \dots , i+k-1]$.

\subsection{Commuting transfer matrices and conserved charges}\label{sec:commuting} Given the set of matrices \\$\{R(u,v)| u,v \in \mathbb{R}\} \subseteq \mathrm{End}(V\otimes V)$, we define the \emph{monodromy matrix} $M_0(u,\xi) \in \mathrm{End} (V\otimes V^{\otimes \ell})$ for every $u\in \mathbb{R}$ and $\xi = (\xi_1, \xi_2 , \dots \xi_\ell) \in \mathbb{C}^\ell$ as
\begin{align*}
    M_0(u,\xi) = \prod_{j=\ell}^1 R_{0j} (u,\xi_j) \; .
\end{align*}
Here we label the tensor components from $0$ to $\ell$, and the tensor component labeled by $0$ has a special role (as all operators in the product act on it); the label $0$ of $M_0(u,\xi)$ emphasizes this. Writing $R(u,\xi) = \sum_i x_i(u,\xi) \otimes y_i(u,\xi)$, with $x_i(u,\xi),y_i(u,\xi)\in \mathrm{End}(V)$, and keeping track of the tensor components explicitly, this expression reads
\begin{align}
    M_0(u,\xi) = \sum_{i_1 ,\dots, i_\ell} \left( x_{i_\ell}(u,\xi_\ell) \cdot x_{i_{\ell-1}}(u,\xi_{\ell-1}) \cdots x_{i_1}(u,\xi_1) \right)  \otimes y_{i_1}(u,\xi_1)  \otimes y_{i_2}(u,\xi_2)  \otimes \dots \otimes y_{i_\ell}(u,\xi_\ell) \; , 
\end{align}
or, writing out the matrix components, 
\begin{align}
    M_0(u,\xi) = \sum_{\bm{\alpha}, \bm{\beta}, \bm{\delta}} 
    R_{\alpha_{1} \beta_\ell}^{\alpha_2 \delta_\ell}(u,\xi_1)
    R_{\alpha_2 \beta_{\ell-1}}^{\alpha_3 \delta_{\ell-1}}(u,\xi_2)  \dots 
    R_{\alpha_{\ell} \beta_1}^{\alpha_{\ell+1} \delta_1}(u,\xi_\ell) \,
    \ket{\alpha_1}\bra{\alpha_{\ell+1}} \otimes 
    \ket{\delta_1 \dots \delta_\ell }\bra{\beta_1 \dots \beta_\ell}.
\end{align}
This operator is an MPO with open boundaries. Throughout the paper we use a graphical notation of tensor networks to depict such operators. The notation is explained in \cref{fig:MPO}. Using this notation, $M_0(u,\xi)$ can be depicted as 
\begin{align}
    M_0(u,\xi) = 
    \begin{tikzpicture}[baseline = -1mm, yscale=0.6, xscale = 2.2]
        \path[draw,%
          decoration={%
            markings,%
            mark=at position 0.5  cm with \arrow{triangle 90},%
            mark=at position 2.7  cm with \arrow{triangle 90},%
            mark=at position 5  cm with \arrow{triangle 90},%
            mark=at position 7  cm with \arrow{triangle 90},%
            mark=at position 9.3  cm with \arrow{triangle 90},%
          },%
          postaction=decorate] (-0.7,0) -- (3.7,0);
        \foreach \x/\l in {0/1,1/2,3/\ell} {
            \draw (\x,-1) --++ (0,2);
            \node[tensor] at (\x,0) {$R(u,\xi_\l)$};
        }
        \node[fill=white] at (2,0) {$\dots$};
        \draw[decoration={brace, raise=1mm,amplitude=1mm}, decorate] (-0.1, 1) -- (3.1,1) node[midway, anchor=south, yshift=1.5mm] {$\ell$};
    \end{tikzpicture} \ ,
\end{align}
where the arrows on the horizontal line (the very first tensor component of $M_0$, labeled by $0$) emphasize that the linear operators act there from the left to right: each $R$-matrix is a linear operator on $V\otimes V$, and the input indices are the bottom (maps to the top index) and the left (maps to the right index). That is,
\begin{align}
    R_{\alpha,\beta}^{\gamma,\delta}(u,\xi) = 
    \begin{tikzpicture}[baseline = -1mm, yscale=0.8, xscale = 2.2]
            \draw (-0.7,0) --++ (1.4,0);
            \draw (0,-1) --++ (0,2);
            \node[tensor] at (0,0) {$R(u,\xi)$};
            \node[anchor=south] at (-0.5,0) {$\alpha$};
            \node[anchor=south] at (0.5,0) {$\gamma$};
            \node[anchor=east] at (0,-0.8) {$\beta$};
            \node[anchor=east] at (0,0.8) {$\delta$};
    \end{tikzpicture} \ .
\end{align}

The \emph{transfer matrix} $T(u,\xi)\in \mathrm{End}\big(V^{\otimes \ell}\big)$ with parameters $\xi = (\xi_1, \xi_2 , \dots \xi_\ell) \in \mathbb{C}^\ell$ is then defined as the trace over the first tensor component of this operator (the one labeled by $0$). The resulting operator acts on the Hilbert space $V^{\otimes \ell}$ and is given by
\begin{align}
    T(u,\xi) = \mathrm{Tr}_0 \left( M_0(u,\xi ) \right) = \mathrm{Tr}_0 \left( \prod_{j=\ell}^1 R_{0j} (u,\xi_j) \right) \; . \label{eq:transfer1}
\end{align}
Just as above, we can express this operator as 
\begin{align}
    T(u,\xi) &= \sum_{i_1\dots i_\ell} \mathrm{Tr}\,\big(x_{i_\ell}(u,\xi_\ell) \cdot x_{i_{\ell-1}}(u,\xi_{\ell-1}) \cdots x_{i_1}(u,\xi_\ell)\big)  \cdot y_{i_1}(u,\xi_1)  \otimes y_{i_2}(u,\xi_2)  \otimes \dots \otimes y_{i_\ell}(u,\xi_\ell), \nonumber\\
      &= \sum_{\bm{\alpha}, \bm{\beta}, \bm{\delta}} 
        R_{\alpha_{1} \beta_\ell}^{\alpha_2 \delta_\ell}(u,\xi_1)
        R_{\alpha_2 \beta_{\ell-1}}^{\alpha_3 \delta_{\ell-1}}(u,\xi_2)  \dots 
        R_{\alpha_{\ell} \beta_1}^{\alpha_{1} \delta_1}(u,\xi_\ell) \,
        \ket{\delta_1 \dots \delta_\ell }\bra{\beta_1 \dots \beta_\ell} \; .
\end{align}

The operator $T(u,\xi)$ has the defining form of an MPO with periodic boundary conditions (as in \cref{fig:MPO}, $W_0 = \dots W_n = V_0$ and $V_i' = V_i$), 
\begin{align}
    T(u,\xi) = 
    \begin{tikzpicture}[baseline = -1mm, yscale=0.6, xscale = 2]
        \draw (-0.7,0) rectangle (3.7, -1.2);
        \foreach \x/\l in {0/1,1/2,3/\ell} {
            \draw (\x,-1) --++ (0,2);
            \node[tensor] at (\x,0) {$R(u,\xi_\l)$};
        }
        \node[fill=white] at (2,0) {$\dots$};
        \draw[decoration={brace, raise=1mm,amplitude=1mm}, decorate] (-0.1, 1) -- (3.1,1) node[midway, anchor=south, yshift=1.5mm] {$\ell$}; 
    \end{tikzpicture} \ .
\end{align}
From the YBE, we have
\begin{align}
    R_{ab} (u,v) M_a (u,\xi) M_b (v,\xi) = M_b (v,\xi ) M_a (u,\xi ) R_{ab} (u,v) \; .
\label{eq:RMM1}
\end{align}
We assume that generically $R_{ab}(u,v)$ is invertible and so, taking the partial trace, the transfer matrices commute with each other:
\begin{align}
    \left[ T(u,\xi ) , T(v,\xi ) \right] = 0 \; .
\end{align}
The logarithm of the transfer matrix can be used to generate local conserved quantities \cite{Korepin97,Gaudin_2014,Jones03}.  In many integrable models, including the XXZ chain, the $R$-matrix is normalized so that $R(0)=SW$ (and $\check{R}=\mathbb{I}$). In such a case, taking a homogeneous limit $\xi_n\rightarrow 0$, we have that $Q_n$, defined by
\begin{align}
Q_n = \left(\frac{\rmd}{\rmd u}\right)^{n-1} \log T(u) \Big \vert_{u=0}\ ,
\end{align}
are mutually commuting charges, with $Q_2$ being the Hamiltonian of the integrable model and where $T(u) := \lim_{\xi_n \to 0} T(u,\xi)$.

More generally, local tensors of the MPO monodromy matrix are referred to as Lax operators and denoted by $L(u)$ (with $\xi$ dependence suppressed). In \cref{eq:transfer1} we have $L(u)=R (u,\xi_j)$, i.e.  we simply identify two indices of the $R$-matrix to correspond to the physical indices. Models with transfer matrices constructed in this way are called fundamental integrable models. More generally Lax operators are intertwined by an $R$-matrix in the {RLL} relation:
\begin{align}
\includegraphics[]{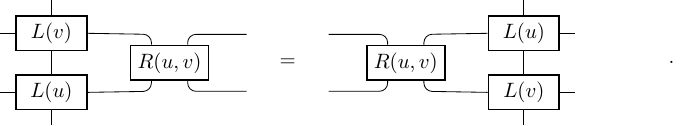}\label{eq:RLLgeneral}
\end{align}
One way to arrive at solutions to this equation, beyond fundamental integrable models, is to use the fusion relations of the $R$-matrix~\cite{Kulish_1981}. One can construct Lax operators with tensored representations, for example the higher-spin representations in the $U_q(\hat{\mathfrak{sl}}_2)$ case\footnote{The XXZ $R$-matrix is related to the spin-$1/2$ evaluation representation of $U_q(\hat{\mathfrak{sl}}_2)$~\cite{charipressley}.}, sharing the same intertwiner $R$. In this article we focus instead on models associated with the fundamental representation, and their MPO duals. 

Assuming that $R$ is invertible, the RLL relation allows us to prove the commutativity of the transfer matrices exactly as above: one inserts $RR^{-1}$ on any bond, pulls $R$ through all Lax operators reversing their order, and then the remaining $R^{-1}R$ can be removed. This is called the `train argument'. In fundamental integrable models, keeping track of indices, the YBE \emph{is} the {RLL} relation. The RLL relation is, however, more general than the YBE.

We remark that mathematically the solutions to the YBE \eqref{eq:YBE_123} can be classified and obtained systematically from the representation theory of quantum groups~\cite{Kulish_1981, charipressley}. We will not focus on this point in the rest of the paper.

\subsection{Baxterization}
Alternatively to the above, Yang--Baxter integrable models can be constructed through a procedure called baxterization. This is summarized in Jones' seminal work \cite{Jones90}, we review the key points here. 

The framework of baxterization is established in purely algebraic terms. Let us consider a unital associative algebra $\mathcal{A}$ over the field $\mathbb{C}$. We assume that there are elements of the algebra $\mathcal{A}$,
\begin{align}
    \{ x_j^{(m)} \} \subset \mathcal{A}\, , \quad j\in \{1,2, \dots, K  \} \, , \quad m \in \{1,2, \dots , n\} \; ,
\end{align}
satisfying certain relations. The aim of the baxterization procedure is to find algebraic solutions, $\check{\mathcal{R}}$, to the circuit YBE
\begin{align}\label{eq:check_R_YBE}
    \check{\mathcal{R}}_{j+1} (u, v) \check{\mathcal{R}}_{j} (u, \xi) \check{\mathcal{R}}_{j+1} (v, \xi) = \check{\mathcal{R}}_{j} (v, \xi)\check{\mathcal{R}}_{j+1} (u, \xi) \check{\mathcal{R}}_{j} (u, v) \; ,
\end{align}
where 
\begin{align}
    \check{\mathcal{R}}_j (u,v) = \sum_{m=1}^n f_n(u,v) x^{(n)}_j \; ,
\end{align}
with certain complex functions $f_n(u,v)$.

There is no assumption on the exact form of the $\check{\mathcal{R}}$, nor the space it acts upon\footnote{We use the terminology $\check{\mathcal{R}}$-matrix to stand for the \emph{abstract} $\check{\mathcal{R}}$ operator.}. The circuit YBE for the $\check{R}$-matrix, given in \cref{eq:checkYBE}, gives concrete Hilbert space solutions (derived from standard $R$-matrices) to the more abstract \cref{eq:check_R_YBE}.
Moreover, \cref{fig:RcheckYBE} is a specific realization of \eqref{eq:check_R_YBE} when the $\check{\mathcal{R}}$-matrix acts on two neighboring sites of a tensor product Hilbert space. 

In order to construct the transfer matrix in this approach, we further require that
\begin{align}
    \left[ \check{\mathcal{R}}_j (u_1,v_1) , \check{\mathcal{R}}_k (u_2,v_2) \right] = 0 , \quad \quad |j-k| \geq 2 \; .
\end{align}
Then we are able to define the \emph{baxterized} transfer matrix following \cite{Jones90},
\begin{align}
    \mathcal{T}(u,\xi) = \check{\mathcal{R}}_K (u,\xi) \check{\mathcal{R}}_{K-1} (u,\xi) \cdots \check{\mathcal{R}}_{2} (u,\xi) \check{\mathcal{R}}_{1} (u,\xi) \; .\label{eq:baxterizedT}
\end{align}
By recursively applying the YBE \eqref{eq:check_R_YBE} to the product of two transfer matrices, we have
\begin{align}
    \check{\mathcal{R}}_K(v,\xi)^{} \check{\mathcal{R}}_K(u,v)^{} \check{\mathcal{R}}_K^{-1} (u,\xi) \mathcal{T}(u,\xi ) \mathcal{T}(v,\xi )  = \mathcal{T}(v,\xi) \mathcal{T} (u,\xi)  \check{\mathcal{R}}_1^{-1} (u,\xi) \check{\mathcal{R}}_1 (u,v)^{} \check{\mathcal{R}}_1^{}(v,\xi) \; .
\end{align}
``Thus up to boundary terms the transfer matrices commute''~\cite{Jones90}. In many examples, it is possible to construct truly commuting transfer matrices from the baxterized transfer matrices, dealing more carefully with the boundary terms.
For cases where $\check{\mathcal{R}} \mapsto \check{R}$, derived from a standard $R$-matrix, then the baxterized transfer matrices \eqref{eq:baxterizedT} give rise\footnote{With appropriate simple manipulations, as they are not strictly equal.} to the standard transfer matrix \eqref{eq:transfer1}.

As for the solutions to \eqref{eq:check_R_YBE}, there are two renowned examples: 1.~solutions with the Temperley--Lieb~(TL) algebra~\cite{TL_1971} and 2.~solutions with the Birman--Murakami--Wenzl~(BMW) algebra~\cite{Birman_1989, Murakami_1987}. By taking different representations of the Hecke algebra \cite{Pasquier_1990, Batchelor_1990, Levy_1991, Nichols_2006, Aufgebauer_2010}, for example, in the TL case, we have the 6-vertex model (XXZ spin chain), the solid-on-solid~(SOS) model, the alternating fundamental--anti-fundamental $SU(n)$ chain, $N$-state Potts model and more. We thus obtain a variety of different models, whose integrable structures are all described by the TL algebra. 
We consider here the TL algebra defined in a field $\mathbb{C}[\beta]$ with $\beta = q + q^{-1}$, where the parameter $q\in \mathbb{C}^{\times}$. When the fugacity parameters $\beta$ of two TL algebras coincide, the spectra of the quantum Hamiltonian as the sum of all TL generators share a similar structure, though the degeneracies might differ~\cite{Alcaraz_1987}. Some of them are more explicitly related by dualities, such as the 6-vertex model and the SOS model, where certain eigenvectors can be mapped to each other through the duality transformation~\cite{Baxter:1972wf, Kojima:2005qz, Ikhlef:2016sgm}. Therefore, the approach of baxterization gives us a powerful way to study integrable models that share the same algebraic structure.

We focus here on the case of chromatic algebra with $3$-colouring~\cite{Fendley_2009, Fendley_2010, Eck24}, which is closely related to the $SO(3)$ BMW algebra at root of unity $q=e^{\rmi \pi /3}$.
This chromatic algebra with $3$-colouring is defined by the generators $\{1,S_j, P_j\}$ with $j \in \{ 1,2, \cdots , K\}$. They satisfy the following relations~\cite{Eck24}
\begin{align}
    S_j^2 = 1 - P_j \, , \quad P_j^2 = P_j \, , \quad S_j P_j = 0 \, , \quad S_j S_{j \pm 1} S_j = P_j S_{j\pm 1} P_j = 0 \; . 
    \label{eq:relationschromatic}
\end{align}
Additional relations can be derived from the ones above. Meanwhile, a `periodic boundary condition' is assumed, i.e. $S_{K+1} = S_1$ and $P_{K+1} = P_1$.
Using the generators of the chromatic algebra, we define the $\check{\mathcal{R}}$-matrix as
\begin{align}
    \check{\mathcal{R}}_j (u;\mu) = \mathbb{I} + \frac{\sin u}{\sin \mu} S_j + \frac{\sin(u+\mu)-\sin \mu}{\sin \mu} P_j \; ,
\end{align}
where $\mu \in \mathbb{C}$ is a fixed parameter.

Using the relations \eqref{eq:relationschromatic}, the $\check{\mathcal{R}}$-matrix satisfies the YBE~\eqref{eq:check_R_YBE}
\begin{align}
    \check{\mathcal{R}}_{j+1} (u-v;\mu) \check{\mathcal{R}}_{j} (u;\mu) \check{\mathcal{R}}_{j+1} (v;\mu) = \check{\mathcal{R}}_{j} (v;\mu) \check{\mathcal{R}}_{j+1} (u;\mu) \check{\mathcal{R}}_{j} (u-v;\mu) \; ;
\end{align}
that is, for each fixed value $\mu$ these operators provide a solution to the YBE \eqref{eq:check_R_YBE} via setting $\check{\mathcal{R}}(u,v) = \check{\mathcal{R}}(u-v;\mu)$.

As shown in \cite{Eck24}, the $\check{R}$-matrices of the XXZ spin chain (and three other integrable models) form a representation of the chromatic algebra of $3$-colouring, which we will exploit in \cref{sec:KW}.

\section{Dualities between lattice models---high level observations} 
\label{sec:dualities}

In this paper, we focus on different duality transformations over integrable spin chains, where the integrability is preserved but the local structures may be modified. The duality transformations act on the physical indices, as do the Hamiltonian operator and the $\check{R}$-matrices. 

To begin with, we classify the duality transformations into three different categories,
\begin{enumerate}
    \item \emph{Invertible} and \emph{on-site} duality transformations.
    \item \emph{Invertible} and \emph{non-on-site} duality transformations. We restrict this (a priori extremely broad) class to those that can be expressed as translation-invariant MPOs with bond dimension $\chi \geq 2$.
    \item \emph{Non-invertible} duality transformations. Again, we restrict this class to those that can be expressed as MPOs with finite bond dimension. Physically relevant examples are often related to the gauging of a discrete symmetry of the Hamiltonian. 
\end{enumerate}

We consider the case when the Hilbert space of the quantum integrable spin chain is $V_1 \otimes V_2 \otimes \cdots \otimes V_\ell$, where $V_n$, $n \in \{1,2,\dots, \ell \}$ is a finite-dimensional vector space. Translation-invariance, possibly after blocking sites, allows us to vary $\ell$ in a given model and to take the thermodynamic limit.

\subsection{Invertible and on-site duality transformations}
\label{subsec:invertibleonsite}

The invertible and on-site duality transformation can be realized by considering the operator
\begin{align}
    U = \prod_{j=1}^{\ell} \mathbf{u}_j \; , \quad \mathbf{u}_j \mathbf{u}_j^{-1} = 1 \; ,
\end{align}
where $\mathbf{u}_j$ is an invertible local operator\footnote{In principle, $\mathbf{u}_j$ can vary with $j$. However, we focus on the translation-invariant case so that momentum remains a conserved quantity and the dependence on $\ell$ is straightforward. } acting non-trivially on $V_j$ only. 

We define the transformed monodromy matrix and the transformed transfer matrix as
\begin{align}
    \tilde{M}_0 (u, \xi ) = U M_0 (u,\xi ) U^{-1} , \quad \tilde{T}(u, \xi ) = \mathrm{Tr}_0 \left(  U M_0 (u,\xi ) U^{-1} \right) = U T(u, \xi ) U^{-1} \; .
\end{align}
It is straightforward to observe that the relation \eqref{eq:RMM1} remains the same,
\begin{align}
    R_{ab} (u,v) \tilde{M}_a (u, \xi ) \tilde{M}_b (v,\xi ) = \tilde{M}_b (v,\xi ) \tilde{M}_a (u,\xi ) R_{ab} (u,v) \; .
\label{eq:RMM2}
\end{align}
Therefore, the transformed transfer matrices are also mutually commuting for all values of the spectral parameters,
\begin{align}
    \left[ \tilde{T}(u, \xi ) , \tilde{T}(v, \xi ) \right] = 0 \; .
\end{align}

The integrable structure remains the same, since the $R$-matrix that intertwines two monodromy matrices is the same. Naturally the $\check{R}$-matrix continues to satisfy the circuit YBE.

\subsection{Invertible with MPO inverse and not-on-site MPO duality transformations}
The second class comprise MPOs (see also \cref{eq:transfer1}) of the form
\begin{align}
    U = \tr_0\left(\prod_{j=\ell}^1  u_{0j}\right) \; , \quad u= \sum_{\alpha,\beta}\ket{\alpha}\bra{\beta} \otimes  (\mathbf{u})_{\alpha,\beta}  \; ,
\end{align}
where $u_{0j}$ is $u$ acting on particles 0 and $j$, and $U$ is such that there exists a second MPO $U^{-1}$, with local tensor $u'_j$, such that $UU^{-1}=\mathbb{I}$ on the physical Hilbert space. When $U$ is unitary, the MPO is an MPU, and the local tensor $ (\mathbf{u'}_j)_{\alpha,\beta} = (\mathbf{u}_j^\dagger)_{\alpha,\beta} $ where we take the complex conjugate and transpose the physical indices (the corresponding tensor is denoted $\overline{u}_j$ in the graphical notation). 

In general, non-on-site transformations do not generate interesting dualities since they will transform local conserved charges to non-local operators. However, for any invertible MPO with MPO inverse, we have the result that local operators are transformed to local operators \cite{Rubio24}, see \cref{sec:locality}. (One manifestation of this is the result that translation-invariant MPU are equivalent to quantum cellular automata \cite{Cirac17,Sahinoglu18}.) This means that the transformed model remains physical. Moreover, if we are interested in Hermitian Hamiltonians, it is natural to focus on the MPU case. This is explored in \cref{sec:cluster}. However, our key observations hold also in the broader family specified. Note also that this class of dualities preserves the spectrum of the integrable model.

Because MPO transformations, ${U}$, with MPO inverse preserve locality, the integrable structure of the $\check{R}$-matrix remains the same:
\begin{align}
    \check{\mathtt{R}}_j (u,v) \check{\mathtt{R}}_{j+1} (u,\xi) \check{\mathtt{R}}_j (v,\xi) = \check{\mathtt{R}}_{j+1} (v,\xi) \check{\mathtt{R}}_j (u,\xi) \check{\mathtt{R}}_{j+1} (u,v) \; ,
\end{align}
where
\begin{align}
    \check{\mathtt{R}}_j (u,v) = U \check{R}_j(u,v) U^{-1} \; .
\end{align}

In contrast, the integrable structure in terms of the $R$-matrix is dramatically modified. Indeed, while we maintain a family of mutually commuting transfer matrices, we no longer expect to be able to find an $R$-matrix satisfying the YBE. However, the transfer matrices in the dual model are defined in terms of local tensors $\tilde{L}$, and the local structure of the MPO transformation allows us to define a modified RLL relation. To do this we construct an operator $\tilde{R}$ from the original $R$-matrix and from the duality transformation (see \cref{extendedR} for the definition, note this is \emph{not invertible} and outside the usual classification of $R$-matrices). Unlike in the standard RLL relation, the combination 
\begin{align}
\includegraphics[]{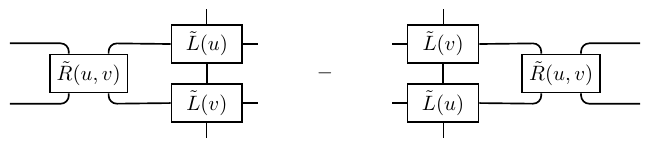} 
\end{align}
is not vanishing. Nevertheless, this expression is orthogonal to the projector $P$ that projects onto the range of $\tilde R$ and this allows us to intertwine the $\tilde{L}$-operators:
\begin{align}
    \begin{tikzpicture}[baseline=-1mm, xscale=2,yscale=0.5]
        \draw (1,-2.3) -- (1,2.3);
        \draw[thick, rounded corners] (0.2,1)  rectangle (1.8,-1);
        \draw[thick, rounded corners] (-0.7,1) -- (-0.2,1) -- (-0.2,-1) -- (-0.7,-1);
        \draw[thick, rounded corners] (2.7,1) -- (2.2,1) -- (2.2,-1) -- (2.7,-1);
        \node[draw,thick,fill=white] at (1,1) {$\tilde{L}(u)$};
        \node[draw,thick,fill=white] at (1,-1) {$\tilde{L}(v)$};
        \node[draw,thick,fill=white] at (0,0) {$\tilde{R}(u,v)$};
        \node[draw,thick,fill=white,minimum width=1.3cm] at (2,0) {$P$};
    \end{tikzpicture} -
    \begin{tikzpicture}[baseline=-1mm, xscale=-2,yscale=-0.5]
        \draw (1,-2.3) -- (1,2.3);
        \draw[thick, rounded corners] (0.2,1)  rectangle (1.8,-1);
        \draw[thick, rounded corners] (-0.7,1) -- (-0.2,1) -- (-0.2,-1) -- (-0.7,-1);
        \draw[thick, rounded corners] (2.7,1) -- (2.2,1) -- (2.2,-1) -- (2.7,-1);
        \node[draw,thick,fill=white] at (1,1) {$\tilde{L}(u)$};
        \node[draw,thick,fill=white] at (1,-1) {$\tilde{L}(v)$};
        \node[draw,thick,fill=white] at (0,0) {$\tilde{R}(u,v)$};
        \node[draw,thick,fill=white,minimum width=1.3cm] at (2,0) {$P$};
    \end{tikzpicture} 
    = 0\ .
\end{align}
We call this the \emph{modified RLL relation}.
In \cref{sec:cluster} we will show that this equation allows us to make the standard train argument for commutativity of transfer matrices, as described in \cref{sec:commuting}. Thus this equation is the relevant local structure underlying the global commutativity of the transfer matrices. Finally, we note that this equation also allows us to recover the modified Yang-Baxter algebra (YBA), as originally defined in Ref.~\cite{Jones22}:
\begin{align}
\includegraphics[]{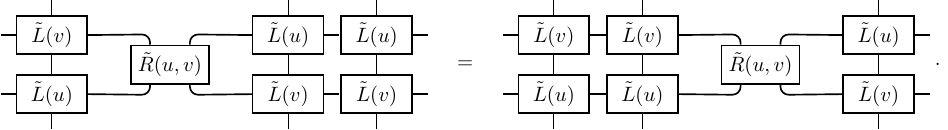} \label{eq:modifiedYBAsec3}
\end{align}

\subsection{Non-invertible MPO duality transformations}\label{sec:noninv}
Non-invertible duality transformations, even when given by exact MPOs, do not necessarily preserve the locality of general local operators~\cite{Jones22}. Here we focus on non-invertible duality transformations that correspond to the gauging of certain discrete symmetries of the Hamiltonian (and $\check{R}$-matrix). Symmetric operators remain local after gauging, and in particular, the Hamiltonian and $\check{R}$-matrix remain local after the duality transformation. This is crucial for integrability.

Suppose that the non-invertible duality transformation is realized by the operator $\mathcal{D}$. We have the following relation,
\begin{align}
    \mathcal{D} \check{R}_j (u,v) = \check{\mathtt{R}}_j (u,v) \mathcal{D} \; ,
\end{align}
where the $\check{R}$-matrix of the dual model is local and given by $\check{\mathtt{R}}_j (u,v)$. The algebraic relations between $\check{R}_j (u,v)$ remain the same for the dual $\check{R}$-matrix, hence the circuit-type Yang--Baxter equations are still satisfied 
\begin{align}
    \check{\mathtt{R}}_j (u,v) \check{\mathtt{R}}_{j+1} (u,w) \check{\mathtt{R}}_j (v,w) = \check{\mathtt{R}}_{j+1} (v,w) \check{\mathtt{R}}_j (u,w) \check{\mathtt{R}}_{j+1} (u,v) \; .
\end{align}
In another word, Jones' baxterization applies for the dual model as well!

It is not obvious how the transfer matrix of the dual model can be constructed, since the operator $\mathcal{D}$ is \emph{non-invertible}. Similarly to the boundary terms in baxterization, the boundary conditions of the transfer matrix must be compatible with the non-invertible duality transformation, i.e. the transfer matrix with the desired boundary condition must be symmetric under the discrete symmetry~\cite{Lootens_2024} (these are known as \emph{topological} boundaries).  In general, it is not guaranteed that an integrable transfer matrix for the dual model exists, especially in the presence of non-trivial topological boundaries. 

In \cref{sec:conclusion} we expand upon the observations summarized here. Before doing so, we study concrete examples of the duality transformations introduced above that illustrate the general features. For this purpose, it is helpful to have a particular integrable model in mind. In \cref{sec:cluster,sec:KW} we demonstrate different duality transformations over a fundamental quantum integrable model, the spin-$\frac{1}{2}$ XXZ spin chain, also known as the 6-vertex model in the statistical mechanics literature. The purpose of the following short section is to fix the notation and clarify the different integrable structures appearing in this model.

\section{The spin-\texorpdfstring{$\frac{1}{2}$}{1/2} XXZ spin chain}
\label{sec:XXZspinchain}

We start with defining the $R$-matrix of the XXZ spin chain (6-vertex model)
\begin{align}
    \mathbf{R} (u;\mu) = \begin{pmatrix} \sin(u+\mu) & {\color{gray}.} & {\color{gray}.} & {\color{gray}.} \\ {\color{gray}.} & \sin u  & \sin \mu & {\color{gray}.} \\ {\color{gray}.} & \sin \mu & \sin u & {\color{gray}.} \\ {\color{gray}.} & {\color{gray}.} & {\color{gray}.} & \sin (u+\mu) \end{pmatrix} \; , \label{eq:RXXZ}
\end{align}
which satisfies the YBE
\begin{align}
    \mathbf{R}_{12} (u-v;\mu) \mathbf{R}_{13} (u;\mu) \mathbf{R}_{23} (v;\mu) = \mathbf{R}_{23} (v;\mu) \mathbf{R}_{13} (u;\mu)   \mathbf{R}_{12} (u-v;\mu) \; ,
\end{align}
where the operator defined on the two sides of this equation acts on three tensored two-dimensional Hilbert spaces labeled by $1,2,3$, $V_1 \otimes V_2 \otimes V_3 = (\mathbb{C}^2)^{\otimes3}$.

The parameter $\mu$ should be considered as a fixed parameter of the model, which is essentially the anisotropy parameter of the XXZ spin chain
\begin{align}
    \Delta = \cos \mu \; ,
\end{align}
while the spectral parameter $u \in \mathbb{C}$ can be chosen to take any value. The $\check{R}$-matrix can be obtained by acting the permutation operator (or swap) of two spin-$\frac{1}{2}$s on the $R$-matrix, i.e.
\begin{align}
    \check{\mathbf{R}}_{12} (u;\mu) = \mathbf{SW}_{12} \mathbf{R}_{12} (u;\mu) = \begin{pmatrix} \sin(u+\mu) & {\color{gray} \cdot } & {\color{gray} \cdot } & {\color{gray} \cdot } \\ {\color{gray}\cdot } & \sin \mu  & \sin u & {\color{gray}\cdot } \\ {\color{gray}\cdot } & \sin u & \sin \mu & {\color{gray}\cdot } \\ {\color{gray}\cdot } & {\color{gray}\cdot } & {\color{gray}\cdot } & \sin (u+\mu) \end{pmatrix} \; ,
\end{align}
where the permutation (swap) operator reads
\begin{align}
    \mathbf{SW} = \begin{pmatrix}
        1 & {\color{gray} \cdot}  & {\color{gray} \cdot}  & {\color{gray} \cdot} \\  {\color{gray} \cdot}  & {\color{gray} \cdot} & 1  & {\color{gray} \cdot} \\   {\color{gray} \cdot} & 1  & {\color{gray} \cdot}  & {\color{gray} \cdot} \\  {\color{gray} \cdot}  & {\color{gray} \cdot} & {\color{gray} \cdot} & 1 
    \end{pmatrix} \; . \label{eq:SW}
\end{align}
The $\check{R}$-matrix satisfies the circuit-type YBE~\eqref{eq:check_R_YBE},
\begin{align}
    \check{\mathbf{R}}_{12} (u-v;\mu) \check{\mathbf{R}}_{23} (u;\mu) \check{\mathbf{R}}_{12} (v;\mu) = \check{\mathbf{R}}_{23} (v;\mu) \check{\mathbf{R}}_{12} (u;\mu)   \check{\mathbf{R}}_{23} (u-v;\mu) \; . 
\end{align}

The transfer matrix of the XXZ spin chain (6-vertex model) is a bond-dimension-2 MPO, i.e.
\begin{align}
    T_{\rm XXZ} (u,\xi ; \mu) = \mathrm{Tr}_0 \mathbf{M}_0 (u,\xi ; \mu) \; , \quad \mathbf{M}_0 (u,\xi ; \mu) = \left[ \prod_{j=\ell}^1 \mathbf{R}_{0j} (u-\xi_j;\mu) \right] \; ,
\end{align}
acting on the Hilbert space $V_1 \otimes V_2 \otimes \cdots \otimes V_\ell = (\mathbb{C}^2)^{\otimes \ell}$, after taking the partial trace over the auxiliary space $V_0 = \mathbb{C}^2$. 
Here we introduce the inhomogeneities as $\xi = (\xi_1, \xi_2 , \dots , \xi_\ell) \in \mathbb{C}^\ell$. The operator $\mathbf{M}_0 \in \mathrm{End} (V_0 \otimes  (\mathbb{C}^2)^{\otimes \ell})$ acting on both the auxiliary 2-dimensional vector space and physical Hilbert space is called the monodromy matrix. By using the YBE recursively, we have
\begin{align}
    \mathbf{R}_{ab} (u-v ; \mu) \mathbf{M}_a (u,\xi ; \mu) \mathbf{M}_b (v,\xi ; \mu) = \mathbf{M}_b (v,\xi ; \mu) \mathbf{M}_a (u , \xi ; \mu) \mathbf{R}_{ab} (u-v ; \mu) \; .
\end{align}
When $\mathbf{R}_{ab} (u ; \mu)$ is invertible (it is not invertible only for a few individual values of $u$), upon inverting it and taking the partial trace over the auxiliary spaces $a$ and $b$, we have
\begin{align}
    \left[ T_{\rm XXZ} (u, \xi ; \mu)  , T_{\rm XXZ} (v, \xi ; \mu) \right] = 0 \; .
\end{align}
For the rest of the article, we will take the homogeneous limit, i.e. $\xi_j =0$, for all $j \in \{ 1,2, \cdots , \ell\}$. In the inhomogeneous case most of the derivation follows the same procedure as well. We will denote the homogeneous transfer matrix with $\xi_j = 0$ as $T_{\rm XXZ} (u ; \mu)$.

The XXZ spin chain Hamiltonian, Eq.~\eqref{eq:XXZ}, is obtained by taking the logarithmic derivative of the transfer matrix evaluated at $u=0$, i.e.
\begin{align}
\begin{split}
    \mathbf{H}_{\rm XXZ} & = \left. 2\sin \mu \, \partial_u \log T_{\rm XXZ} (u ; \mu) \right|_{u=0} = \sum_{j=1}^L 2\check{\mathbf{R}}_{j,j+1}^\prime (0 ; \mu) \\ 
    & = \sum_{j=1}^{\ell} \left( X_j X_{j+1} + Y_j Y_{j+1} +\Delta (\mathbb{I}+Z_j Z_{j+1} ) \right) = \sum_{j=1}^L \mathbf{h}_{j,j+1} \; ,
\end{split}
\label{eq:XXZHamiltonian}
\end{align}
where the Pauli operators $X_j$, $Y_j$ and $Z_j$ act non-trivially on $V_j$. The anisotropy parameter $\Delta = \cos \mu$ is defined as above.

A crucial point is that the $\check{R}$-matrix and the XXZ Hamiltonian can be expressed in terms of a representation of the chromatic algebra mentioned above, with the generators
\begin{align}
    S_j \mapsto \mathbf{S}_{j} = \frac{1}{2} \left( X_j X_{j+1} + Y_j Y_{j+1} \right) \; , \quad P_j \mapsto \mathbf{P}_j = \frac{1}{2} \left(\mathbb{I}+Z_j Z_{j+1} \right) \; .
    \label{eq:XXZrepchromatic}
\end{align}
Therefore, we rewrite the $\check{R}$-matrix in terms of the chromatic algebra generators as
\begin{align}
    \check{\mathbf{R}}_{j,j+1} (u , \mu) = \sin \mu \, \mathbb{I} + \left( \sin (u+\mu) - \sin \mu \right) \mathbf{P}_j + \sin u \, \mathbf{S}_j \; , \quad \mathbf{h}_{j,j+1} = 2 \left( \mathbf{S}_j + \Delta \mathbf{P}_j \right) \; .
\end{align}

\section{Case study 1: the cluster entangler applied to the XXZ chain}\label{sec:cluster}
In this section we analyse the integrable structure of the Hamiltonian $\mathbf{H}_{\rm ZXXZ}$ that arises from transforming the XXZ chain with the `cluster entangler MPU'. This is a $\chi=2$ unitary MPO denoted $U_c$, where $\chi$ denotes the bond dimension. Surprisingly, even though the model is integrable (see \cref{sec:dualities} for a general discussion), its transfer matrix $T_{\rm ZXXZ}$ (or more precisely, the injective MPO tensor generating it) does not admit a conventional invertible intertwiner satisfying the RLL relation and the YBE. Instead, we find that the `uncompressed' MPO tensor describing $T_{\rm ZXXZ}$ (that is, the MPO tensor with $\chi=8$, composed of three $\chi=2$ MPO tensors: those of $U_c$, $T_{\rm XXZ}$ and $U_c^\dagger$) admits an `extended' $R$-matrix satisfying a modified RLL relation. We show that this extended $R$-matrix and the modified RLL relation is the essential local tensor identity underlying the integrability of $\mathbf{H}_{\rm ZXXZ}$. The construction generalizes to any model obtained via MPU duality from a standard Yang-Baxter integrable model. We note that the dual model $\mathbf{H}_{\rm ZXXZ}$ was considered in Ref.~\cite{Jones22}; here we give a new analysis in terms of explicit tensor networks and recast the results as identities intrinsic to the dual model. We further generalize these results to dualities described by MPOs with MPO-inverse in \cref{sec:conclusion}.

\subsection{The cluster entangler and the dual XXZ model}
Let us consider the SPT-entangler for the cluster spin-chain, given by 
\begin{align}
U^{}_c =U_c^\dagger= \prod_{j=1}^{\ell} CZ_{j,j+1} \qquad\mathrm{where}\qquad  CZ=\left(\begin{array}{cccc}1 & {\color{gray} \cdot } & {\color{gray} \cdot } & {\color{gray} \cdot } \\{\color{gray} \cdot } & 1 & {\color{gray} \cdot } & {\color{gray} \cdot } \\{\color{gray} \cdot } & {\color{gray} \cdot } & 1 & {\color{gray} \cdot } \\{\color{gray} \cdot } & {\color{gray} \cdot } & {\color{gray} \cdot } & -1\end{array}\right) \ .
\end{align}
We take periodic boundary conditions, identifying site indices $j$ with $j+\ell$.
This is a Clifford transformation that acts as $U_c Z_j U_c^\dagger = Z_j$ and $U_c X_j U_c^\dagger = Z_{j-1}X_jZ_{j+1}$. Hence, it transforms the trivial paramagnet $\mathbf{H}_0=-\sum_j X_j$ to $\mathbf{H}_{\mathrm{cluster}}=-\sum_j Z_{j-1}X_jZ_{j+1}$. 

Of course, we can use $U_c$ to transform any integrable model. For definiteness, we consider the XXZ spin chain introduced above, giving
\begin{align}
{\mathbf{H}}_{\mathrm{ZXXZ}}= U^{}_c \mathbf{H}_{\mathrm{XXZ}} U_c^\dagger = \sum_{\mathrm{sites}~j} \Big( Z_{j-1}X_j X_{j+1}Z_{j+2} +Z_{j-1} Y_j Y_{j+1}Z_{j+2} + \Delta (\mathbb{I}+Z_jZ_{j+1})  \Big)      \label{eq:XXZ2} \ .
\end{align}
For $\Delta<1$, this model is at a transition between two distinct symmetry-broken phases \cite{Sachdev11}, where the ground state is also an SPT for the unbroken symmetry group \cite{Verresen17,Rey25}.
In particular, in the absence of an external field, the XXZ chain has symmetries $\prod_j Z_j$, $\prod_j X_j$ along with time-reversal. The latter acts as complex conjugation, $\mathcal{K}$, in the $Z$ basis. Perturbing $\mathbf{H}_{\mathrm{XXZ}}$ with $\delta(X_j X_{j+1}-Y_jY_{j+1})$ takes us to an $X$ or $Y$ ordered antiferromagnetic phase for $\delta>0$ and $\delta<0$ respectively. Applying the entangler, the $X$-antiferromagnet transforms to a state with long-range order in $Z_{j-1}X_jZ_{j+1}$. This implies symmetry fractionalisation of the preserved $\prod_j X_j$ and $\mathcal{K}$ symmetries. This is in line with our motivation of using dualities to construct integrable transitions between SPT phases.

The transformation $U^{}_c$ is itself an MPU with MPU tensors\footnote{From this point on, following the conventions of the MPO literature, we read the virtual indices of the MPO tensors from right to left, and not from left to right.} given by
\begin{align}
    \begin{tikzpicture}[baseline = -1mm, yscale=0.8, xscale = 2.2]
            \draw (-0.7,0) --++ (1.4,0);
            \draw (0,-1) --++ (0,2);
            \node[tensor] at (0,0) {$U_c$};
            \node[anchor=south] at (-0.5,0) {$\alpha$};
            \node[anchor=south] at (0.5,0) {$\gamma$};
            \node[anchor=east] at (0,-0.8) {$\beta$};
            \node[anchor=east] at (0,0.8) {$\delta$};
    \end{tikzpicture} \ = (-1)^{\alpha\gamma} \cdot  \delta_{\alpha \beta \delta}\ .
\end{align}
It is straightforward to check that its square admits the following decomposition\footnote{We use bold lines to indicate the larger bond Hilbert space here.} \cite{Sahinoglu18}:
\begin{align}\label{eq:MPUdecomposition}
\includegraphics[]{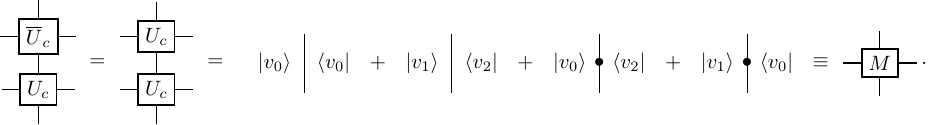}
\end{align}
Here $\bullet$ indicates a $Z$ operator, and we have orthonormal vectors on the virtual space given by
\begin{align}
\ket{v_0} &=\frac{1}{\sqrt{2}} \left(\ket{00}+\ket{11}\right) \nonumber\\
\ket{v_1} &=\frac{1}{\sqrt{2}} \left(\ket{00}-\ket{11}\right) \nonumber\\
\ket{v_2} &=\frac{1}{\sqrt{2}} \left(\ket{01}+\ket{10}\right) \ .
\end{align}
This is a `fixed point form' for the MPU.
In particular, it satisfies the following fixed-point equations
\begin{align}
\includegraphics[]{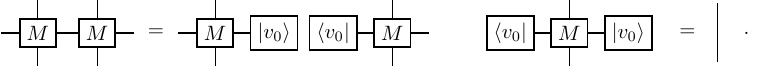}\label{eq:fixedpoint}
\end{align}
These equations can be verified for this particular MPU, but hold more generally for any MPU (after blocking sites, and where $\ket{v_0}$ and $\bra{v_0}$ are replaced by appropriate vectors) \cite{Cirac17,Sahinoglu18}. See also \cref{sec:ABtransfer}. These relations imply the `pulling through' equation
\begin{align}
\includegraphics[]{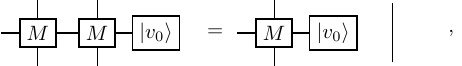}\label{eq:pullingthrough}
\end{align}
and its mirror image.
The transfer matrix $T_{\mathrm{ZXXZ}}(u)=U^{}_cT_{\mathrm{XXZ}}(u)U^{}_c$ for $\mathbf{H}_{\mathrm{ZXXZ}}$ is naturally constructed from Lax operators
$L_{\mathrm{ZXXZ}}(u):=\tilde{L}(u)$ given diagrammatically by
\begin{align}
\includegraphics[]{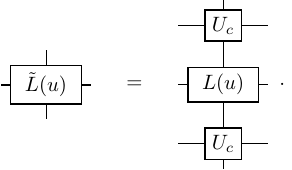} \label{extendedL}
\end{align}
Here $L$ is the Lax operator of the XXZ chain (constructed from $R$ given in \cref{eq:RXXZ}).

Globally the transfer matrices $T_{\mathrm{ZXXZ}}(u)$, constructed from $\tilde{L}(u)$, commute for different values of the spectral parameter. This is clear since they are unitarily equivalent to $T_{\mathrm{XXZ}}$. However, since both $U_c$ and the $T_{\mathrm{XXZ}}(u)$ are MPOs that are built from local tensors, one might hope there is a local relation that gives the commutation of the $T_{\mathrm{ZXXZ}}(u)$. Of course, this is not guaranteed on general grounds. 

A first thing to look for is an \emph{invertible} operator $\tilde{R}(u,v)$ that intertwines $\tilde{L}(u)$, i.e. a solution to the RLL relation \cref{eq:RLLgeneral}. This would be impossible for a trivial reason if our representation \cref{extendedL} is not full rank. By applying the singular-value decomposition, we can indeed compress the $\chi=8$ representation of $\tilde{L}(u)$ to a representation with $\chi=6$. Using simple built-in functions in Mathematica \cite{Mathematica} to solve \cref{eq:RLLgeneral} with the compressed $\tilde{L}(u)$, we see that there is a non-trivial space of solutions. However these solutions are not invertible, and moreover do not satisfy the YBE. 

Due to their non-invertibility, the usual train argument, described in \cref{sec:commuting}, does not apply to show commutativity of the transfer matrices. Nevertheless, in the next subsection we explain how we can define an extended $R$-matrix, following \cite{Jones22}, that despite its non-invertibility provides such a local relation, and can be used to generalize the train argument. The extended $R$-matrix that we discuss here contains minimal elements of the two ingredients of the dual model: the original integrable Hamiltonian and the MPU duality. We will also see that this easily generalizes to MPO with MPO-inverse in \cref{sec:conclusion}.

\subsection{The extended $R$-matrix and modified RLL relation}

We cannot immediately substitute the XXZ $R$-matrix into a local RLL relation for $\tilde{L}$, due to the $U_c$ tensors. However, the pulling through equation is suggestive that projecting onto $\ket{v_0} $ on the virtual space is helpful. Indeed, we can define the extended $R$-matrix, denoted $\tilde{R}$, as follows
\begin{align}
\includegraphics[]{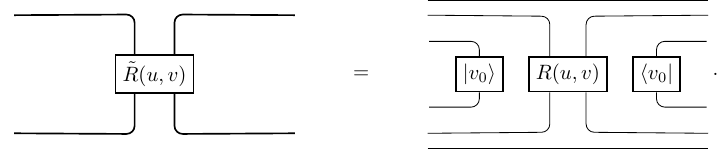} \label{extendedR}
\end{align}
Due to the presence of the projector $\ket{v_0}\bra{v_0}$, this matrix is clearly not invertible: $\tilde{R}(u,v) = \tilde{R}(u,v) P = P \tilde{R}(u,v)$, where 
\begin{align}
 \includegraphics[]{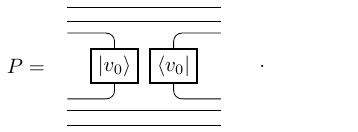}
\end{align}
However, for our purposes we can consider its pseudo-inverse $\tilde{R}^-(u,v)$ given by
\begin{align}
\includegraphics[]{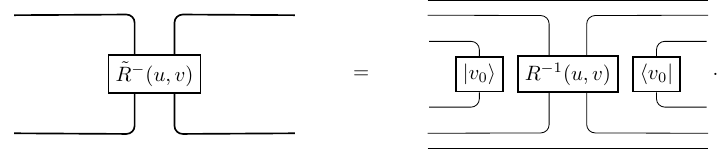} \label{R-}
\end{align}
This matrix $\tilde{R}^-(u,v)$ satisfies the identity $\tilde{R}^-(u,v) \tilde{R}(u,v) = \tilde{R}(u,v)\tilde{R}^-(u,v) = P$, 
\begin{align}
\includegraphics[]{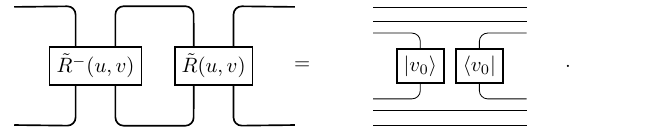} \label{R-R}
\end{align}
While this is not a true inverse, the projector $P$ acts as identity when inserted between the appropriate MPU tensors---see \cref{eq:fixedpoint}:
\begin{equation}\label{eq:LLPLL}
    \begin{tikzpicture}[baseline=-1mm, xscale=2,yscale=0.5]
        \draw (1,-2.3) -- (1,2.3);
        \draw (2,-2.3) -- (2,2.3);
        \draw[thick] (1-0.5,1) -- (2.5,1);
        \draw[thick] (1-0.5,-1) -- (2.5,-1);
        \node[draw,thick,fill=white] at (1,1) {$\tilde{L}(u)$};
        \node[draw,thick,fill=white] at (1,-1) {$\tilde{L}(v)$};
        \node[draw,thick,fill=white] at (2,1) {$\tilde{L}(u)$};
        \node[draw,thick,fill=white] at (2,-1) {$\tilde{L}(v)$};
    \end{tikzpicture} =
    \begin{tikzpicture}[baseline=-1mm, xscale=2,yscale=0.5]
        \draw (1,-2.3) -- (1,2.3);
        \draw (-1,-2.3) -- (-1,2.3);
        \draw[thick, rounded corners] (1.5,1) -- (0.2,1) -- (0.2,-1) -- (1.5,-1);
        \draw[thick, rounded corners] (-1.5,1) -- (-0.2,1) -- (-0.2,-1) -- (-1.5,-1);
        \node[draw,thick,fill=white] at (1,1) {$\tilde{L}(u)$};
        \node[draw,thick,fill=white] at (1,-1) {$\tilde{L}(v)$};
        \node[draw,thick,fill=white,minimum width=1.3cm] at (0,0) {$P$};
        \node[draw,thick,fill=white] at (-1,1) {$\tilde{L}(u)$};
        \node[draw,thick,fill=white] at (-1,-1) {$\tilde{L}(v)$};
    \end{tikzpicture}\ .
\end{equation}
This holds for all pairs $(u,v)$, as $P$ is independent of $u$ and $v$, and in particular, for $(v,u)$ as well.

Moreover, it is straightforward to check that $\tilde{R}(u,v)$ satisfies the YBE \cite{Jones22}. We will now show that $\tilde{R}(u,v)$ is not a true intertwiner, but in spite of this, it can still be used to show that the transfer matrices $T_{ZXXZ}(u)$ and $T_{ZXXZ}(v)$ commute. 

First, we can check that $\tilde{R}(u,v)$ is not a true intertwiner, i.e., it does not satisfy the RLL relations:
\begin{align}
    \begin{tikzpicture}[baseline=-1mm, xscale=2,yscale=0.5]
        \draw (1,-2.3) -- (1,2.3);
        \draw[thick, rounded corners] (1.5,1) -- (0.2,1) -- (0.2,-1) -- (1.5,-1);
        \draw[thick, rounded corners] (-0.7,1) -- (-0.2,1) -- (-0.2,-1) -- (-0.7,-1);
        \node[draw,thick,fill=white] at (1,1) {$\tilde{L}(u)$};
        \node[draw,thick,fill=white] at (1,-1) {$\tilde{L}(v)$};
        \node[draw,thick,fill=white] at (0,0) {$\tilde{R}(u,v)$};
    \end{tikzpicture} -
    \begin{tikzpicture}[baseline=-1mm, xscale=-2,yscale=-0.5]
        \draw (1,-2.3) -- (1,2.3);
        \draw[thick, rounded corners] (1.5,1) -- (0.2,1) -- (0.2,-1) -- (1.5,-1);
        \draw[thick, rounded corners] (-0.7,1) -- (-0.2,1) -- (-0.2,-1) -- (-0.7,-1);
        \node[draw,thick,fill=white] at (1,1) {$\tilde{L}(u)$};
        \node[draw,thick,fill=white] at (1,-1) {$\tilde{L}(v)$};
        \node[draw,thick,fill=white] at (0,0) {$\tilde{R}(u,v)$};
    \end{tikzpicture} \neq 0\ .\end{align}
Indeed, after simple manipulations given in \cref{app:RLL}, the LHS of this equation reduces to
\begin{align}
\includegraphics[]{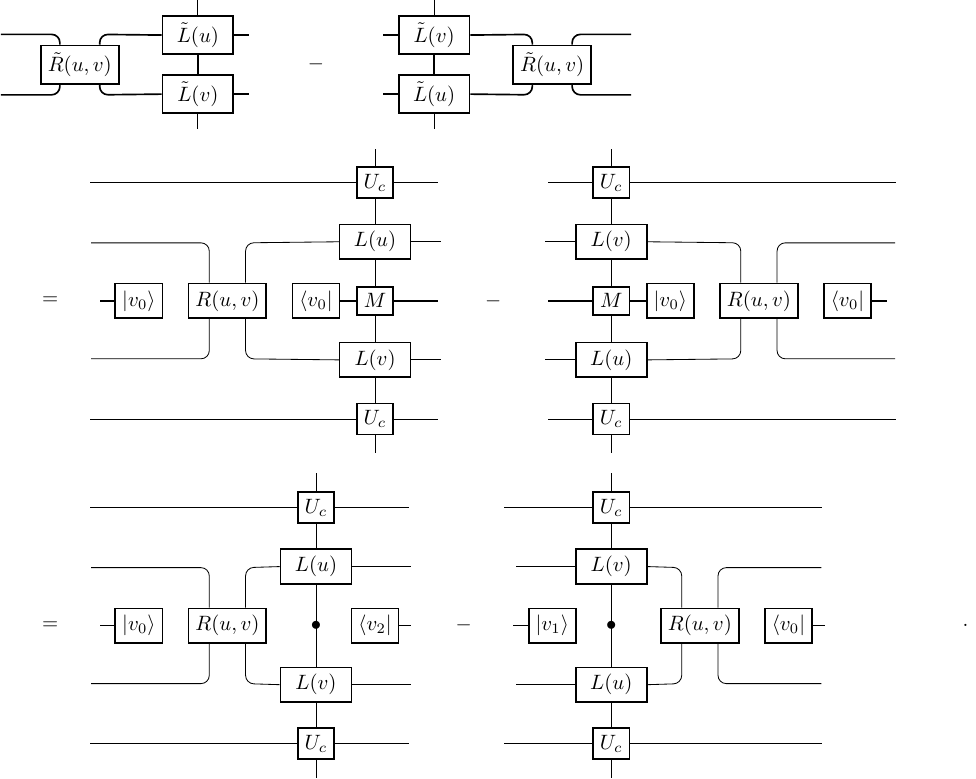} \label{eq:RLL}
\end{align}
In general the right-hand-side is non-vanishing. We confirm this for the particular case studied here in \cref{app:RLL}. 

Let us now show that despite $\tilde{R}(u,v)$ not being an intertwiner, it satisfies local equations (intrinsic to the dual model) that can be used to show that the transfer matrices $T_{ZXXZ}(u)$ and $T_{ZXXZ}(v)$ commute. For that, notice that the right-hand-side of the previous equation, \cref{eq:RLL},  vanishes as soon as we multiply with the projector $P$ defined in \cref{R-R} on its left and right. That is,  as $P \tilde R(u,v) = \tilde R(u,v) P$,
\begin{align}\label{eq:RLLP}
    \begin{tikzpicture}[baseline=-1mm, xscale=2,yscale=0.5]
        \draw (1,-2.3) -- (1,2.3);
        \draw[thick, rounded corners] (0.2,1)  rectangle (1.8,-1);
        \draw[thick, rounded corners] (-0.7,1) -- (-0.2,1) -- (-0.2,-1) -- (-0.7,-1);
        \draw[thick, rounded corners] (2.7,1) -- (2.2,1) -- (2.2,-1) -- (2.7,-1);
        \node[draw,thick,fill=white] at (1,1) {$\tilde{L}(u)$};
        \node[draw,thick,fill=white] at (1,-1) {$\tilde{L}(v)$};
        \node[draw,thick,fill=white] at (0,0) {$\tilde{R}(u,v)$};
        \node[draw,thick,fill=white,minimum width=1.3cm] at (2,0) {$P$};
    \end{tikzpicture} -
    \begin{tikzpicture}[baseline=-1mm, xscale=-2,yscale=-0.5]
        \draw (1,-2.3) -- (1,2.3);
        \draw[thick, rounded corners] (0.2,1)  rectangle (1.8,-1);
        \draw[thick, rounded corners] (-0.7,1) -- (-0.2,1) -- (-0.2,-1) -- (-0.7,-1);
        \draw[thick, rounded corners] (2.7,1) -- (2.2,1) -- (2.2,-1) -- (2.7,-1);
        \node[draw,thick,fill=white] at (1,1) {$\tilde{L}(u)$};
        \node[draw,thick,fill=white] at (1,-1) {$\tilde{L}(v)$};
        \node[draw,thick,fill=white] at (0,0) {$\tilde{R}(u,v)$};
        \node[draw,thick,fill=white,minimum width=1.3cm] at (2,0) {$P$};
    \end{tikzpicture} 
    = 0\ .
\end{align}
 \cref{eq:RLLP} implies the modified YBA of Ref.~\cite{Jones22}: 
\begin{align}
\includegraphics[]{modifiedYBA.pdf} \label{eq:modifiedYBA}
\end{align}
Finally note that \cref{R-R,eq:LLPLL,eq:RLLP} can be used to repeat the train argument presented in \cref{sec:commuting} proving that the transfer matrices $T_{ZXXZ}(u)$ and $T_{ZXXZ}(v)$ commute: First, using \cref{eq:LLPLL} we insert $P$ between every pair of $\tilde{L}$ tensors in $\tilde T(u) \cdot \tilde T(v)$. Second, we write $P = \tilde R^{-}(u,v) \tilde R(u,v)$ (i.e., \cref{R-R}) for one of the $P$s. Third, using \cref{eq:RLLP}, the matrix $\tilde R(u,v)$ slides through the chain exchanging $\tilde L(u)$ and $\tilde L(v)$, and due to the periodic boundaries, it meets $\tilde R^{-}(u,v)$ from the right again, resulting in $\tilde R(u,v) \tilde R^{-}(u,v) = P$. Finally, using again \cref{eq:LLPLL} with $u$ and $v$ exchanged, we can remove all $P$s, resulting in $\tilde T(v) \cdot \tilde T(u)$.

Let us remark that, as the original $R$-matrix depends only on the difference $(u-v)$, $\tilde R(u,v)$ also depends only on $(u-v)$, and thus this argument
also works to show that the transfer matrices of integrable models with inhomogeneities (i.e. different spectral parameters appearing in the $\tilde{L}$ at different sites, cf. \eqref{eq:transfer1}) commute (as long as the difference of the spectral parameters is the same at each site). The example of the XXZ spin chain that we focus on is a uniform case, so we presented this specialisation for clarity.

\section{Case study 2: vertex-face correspondence and Kramers--Wannier}\label{sec:KW}
The non-invertible MPO dualities, as discussed in \cref{sec:noninv}, are motivated by gauging discrete global symmetries~\cite{Bhardwaj:2017xup}, and can be interpreted as maps between different module categories of a fusion category~\cite{Lootens_2023, Lootens_2024}. Because locality is preserved only for symmetry-invariant operators, it is natural in the integrable setting to focus on Hamiltonians and $\check{R}$-matrices carrying the relevant symmetry. Under the duality transformation, such $\check{R}$-matrices transform into duals that continue to satisfy the Yang–Baxter equation. This yields a local dual Hamiltonian with manifest integrable structure.

In this section we make this concrete by focusing on the Kramers–Wannier (KW) duality \cite{Kramers41} between the XXZ spin chain and the Ising zig-zag model. We demonstrate that the KW duality transformation, despite being non-invertible, leaves the integrable structure intact. We construct the transfer matrix $T_{\rm Izz}$ of the Ising zig-zag model as a split-index matrix product operator (that is, a face-type transfer matrix) and show that the KW duality transforms the transfer matrix of the XXZ model into $T_{\rm Izz}$. The integrability of the Ising zig-zag model can thus be demonstrated explicitly, and the KW duality can be interpreted as a vertex-face correspondence. We expect analogous constructions to hold for other non-invertible dualities over integrable spin chains.

We begin with the XXZ Hamiltonian \eqref{eq:XXZHamiltonian},
\begin{align}
    \mathbf{H}_{\rm XXZ} = \sum_{j=1}^{\ell} \left( X_j X_{j+1} + Y_j Y_{j+1} +\Delta (\mathbb{I}+Z_j Z_{j+1} ) \right) \; ,
\end{align}
with $\Delta= \cos \mu$. Its associated $\check R$-matrix in terms of chromatic algebra generators \eqref{eq:XXZrepchromatic} reads
\begin{align}
    \check{\mathbf{R}}_{j,j+1} (u , \mu) = \sin \mu \, \mathbb{I} + \left( \sin (u+\mu) - \sin \mu \right) \mathbf{P}_j + \sin u \, \mathbf{S}_j \; .
\end{align}
The KW MPO gauges the global $\mathbb{Z}_2$ symmetry generated by
\begin{equation}
    \eta = \prod_{j=1}^{\ell} Z_j \; .
\end{equation}
Under this transformation, the XXZ spin chain is mapped to the Ising zig-zag model~\cite{Aasen20,Zadnik21,Pozsgay21,Gombor21,Eck24}. Importantly, the corresponding integrable structure remains local and can be described entirely within the baxterization framework. The dual Ising zig-zag Hamiltonian obtained after the KW transformation is given by
\begin{align}
    \mathbf{H}_{\rm Izz} = \sum_{j=1}^{\ell} Z_{j+1/2} (\mathbb{I} - X_{j-1/2} X_{j+3/2} ) + \Delta (\mathbb{I} + X_{j-1/2} X_{j+3/2} ) \; .
\label{eq:IzzHamiltonian}
\end{align}
As we now show, this Hamiltonian inherits a local Yang--Baxter integrable structure from the XXZ chain.

The KW MPO is a bond-dimension two MPO,
\begin{align}
    \mathcal{D}_{\rm KW} = \mathrm{Tr}_a \prod_{j=1}^{\ell} \left( \mathbf{A}_{a,j-1/2} \mathbf{B}_{a,j} \right) \; , 
\end{align}
which can be represented diagrammatically as
\begin{align}
\raisebox{-.5\height}{
\begin{tikzpicture}[scale=1.0]
\begin{scope}[shift={({0},{0})}]
 \node (node1) at (0,0) [thick,minimum size=.15cm, minimum width = 0.8cm,circle,draw] {$A$};
 \node (node2) at (2,0) [thick,minimum size=.15cm, minimum width = 0.8cm,circle,draw] {$B$};
 \node (node3) at (4,0) [thick,minimum size=.15cm, minimum width = 0.8cm,circle,draw] {$A$};
  \node (node4) at (6,0) [thick,minimum size=.15cm, minimum width = 0.8cm,circle,draw] {$B$};
  \node (node5) at (8,0) [thick,minimum size=.15cm, minimum width = 0.8cm,circle,draw] {$A$};
  \node (node6) at (10,0) [thick,minimum size=.15cm, minimum width = 0.8cm,circle,draw] {$B$};
    \coordinate (h0) at (-1,0) {};
    \coordinate (h1) at (-0.4,0) {};
    \coordinate (h2) at (0.4,0) {};
    \coordinate (h3) at (1.6,0) {};
    \coordinate (h4) at (2.4,0) {};
    \coordinate (h5) at (3.6,0) {};
    \coordinate (h6) at (4.4,0) {};
    \coordinate (h7) at (5.6,0) {};
    \coordinate (h8) at (6.4,0) {};
    \coordinate (h9) at (7.6,0) {};
    \coordinate (h10) at (8.4,0) {};
    \coordinate (h11) at (9.6,0) {};
    \coordinate (h12) at (10.4,0) {};
    \coordinate (h13) at (11,0) {};
    \coordinate (u0) at (0,0.4) {};
    \coordinate (u1) at (0,1.5) {};
    \coordinate (u2) at (4,0.4) {};
    \coordinate (u3) at (4,1.5) {};
    \coordinate (u4) at (8,0.4) {};
    \coordinate (u5) at (8,1.5) {};
    \coordinate (d0) at (2,-0.4) {};
    \coordinate (d1) at (2,-1.5) {};
    \coordinate (d2) at (6,-0.4) {};
    \coordinate (d3) at (6,-1.5) {};
    \coordinate (d4) at (10,-0.4) {};
    \coordinate (d5) at (10,-1.5) {};
    \draw[color=black] (h0) -- (h1);
    \draw[color=black] (h2) -- (h3);
    \draw[color=black] (h4) -- (h5);
    \draw[color=black] (h6) -- (h7);
    \draw[color=black] (h8) -- (h9);
    \draw[color=black] (h10) -- (h11);
    \draw[color=black] (h12) -- (h13);
    \draw[color=black] (u0) -- (u1);
    \draw[color=black] (u2) -- (u3);
    \draw[color=black] (u4) -- (u5);
    \draw[color=black] (d0) -- (d1);
    \draw[color=black] (d2) -- (d3);
    \draw[color=black] (d4) -- (d5);
\end{scope}
\end{tikzpicture}
}\ .
\end{align} 
Notice the shift from sites to bonds \cite{Aasen16}. The local tensors are
\begin{align}
\begin{split}
\raisebox{-.5\height}{
\begin{tikzpicture}[scale=1.0]
\begin{scope}[shift={({0},{0})}]
 \node (node1) at (0,0) [thick,minimum size=.15cm, minimum width = 0.8cm,circle,draw] {$A$};
 \coordinate (h0) at (-1,0) {};
    \coordinate (h1) at (-0.4,0) {};
    \coordinate (h2) at (0.4,0) {};
    \coordinate (h3) at (1,0) {};
    \coordinate (u0) at (0,0.4) {};
    \coordinate (u1) at (0,1) {};
    \draw[color=black] (h0) -- (h1);
    \draw[color=black] (h2) -- (h3);
    \draw[color=black] (u0) -- (u1);
 \end{scope}
 \end{tikzpicture}} & = 
 \begin{pmatrix} |-\rangle_{n-1/2} & |+\rangle_{n-1/2} \\ |-\rangle_{n-1/2} & -|+\rangle_{n-1/2} \end{pmatrix}_a \\
 & =  \frac{1}{\sqrt{2}}\begin{pmatrix} 1 & 1 \\ 1 & -1 \end{pmatrix}_a | 0 \rangle_{n-1/2} + \frac{1}{\sqrt{2}}\begin{pmatrix} -1 & 1 \\ -1 & -1 \end{pmatrix}_a | 1 \rangle_{n-1/2} \; ,
\end{split}
\end{align}
and 
\begin{align}
\begin{split}
\raisebox{-.5\height}{
\begin{tikzpicture}[scale=1.0]
\begin{scope}[shift={({0},{0})}]
 \node (node1) at (0,0) [thick,minimum size=.15cm, minimum width = 0.8cm,circle,draw] {$B$};
 \coordinate (h0) at (-1,0) {};
    \coordinate (h1) at (-0.4,0) {};
    \coordinate (h2) at (0.4,0) {};
    \coordinate (h3) at (1,0) {};
    \coordinate (d0) at (0,-0.4) {};
    \coordinate (d1) at (0,-1) {};
    \draw[color=black] (h0) -- (h1);
    \draw[color=black] (h2) -- (h3);
    \draw[color=black] (d0) -- (d1);
 \end{scope}
 \end{tikzpicture}} & = \begin{pmatrix} \langle - |_{n} & \langle + |_{n} \\ \langle-|_{n} & - \langle + |_{n} \end{pmatrix}_a \\
 & =  \frac{1}{\sqrt{2}}\begin{pmatrix} 1 & 1 \\ 1 & -1 \end{pmatrix}_a \langle 0 |_{n} + \frac{1}{\sqrt{2}}\begin{pmatrix} -1 & 1 \\ -1 & -1 \end{pmatrix}_a \langle 1 |_{n} \; .
\end{split}
\end{align}

The KW MPO realizes the gauging of the global $\mathbb{Z}_2$ symmetry $\eta$, which in terms of operators defined on the Hilbert space $(\mathbb{C}^2)^{\otimes \ell}$ means
\begin{align}
\begin{split}
    & \mathcal{D}_{\rm KW}^\dagger \mathcal{D}_{\rm KW}^{\vphantom \dagger} = \mathbb{I} + \prod_{j=1}^\ell Z_j = \mathbb{I} + \eta \; ,  \\
    & \mathcal{D}_{\rm KW}^{\vphantom \dagger} \mathcal{D}_{\rm KW}^\dagger = \mathbb{I} + \prod_{j=1}^\ell Z_{j-1/2} = \mathbb{I} + \tilde{\eta} \; ,
\end{split}
\end{align}
where $\tilde{\eta}$ is the global $\mathbb{Z}_2$ symmetry of the dual model\footnote{In general for a model with global finite group symmetry $G$, the dual model has the global symmetry $\mathrm{Rep} (G)$, i.e. the representation category of $G$. For finite Abelian group, $\mathrm{Rep}(G) \simeq G$ when there is no anomaly. This holds for the KW MPO transformation.}.

The action of the KW MPO on local operators is
\begin{align}
    \mathcal{D}_{\rm KW} Z_n = X_{n-1/2} X_{n+1/2} \mathcal{D}_{\rm KW} \; , \quad \mathcal{D}_{\rm KW} X_n X_{n+1} = Z_{n+1/2} \mathcal{D}_{\rm KW} \; .
\end{align}
These equations are derived from the relation that the local tensors satisfy,
\begin{align}
\raisebox{-.5\height}{
\begin{tikzpicture}[scale=1.0]
\begin{scope}[shift={({0},{0})}]
 \node (node1) at (0,0) [thick,minimum size=.15cm, minimum width = 0.8cm,circle,draw] {$A$};
 \node (node2) at (2,0) [thick,minimum size=.15cm, minimum width = 0.8cm,circle,draw] {$B$};
 \node (node3) at (4,0) [thick,minimum size=.15cm, minimum width = 0.8cm,circle,draw] {$A$};
 \node (node4) at (2,-1.5) [thick,minimum size=.15cm, minimum width = 0.8cm,circle,draw] {$Z$};
    \coordinate (h0) at (-1,0) {};
    \coordinate (h1) at (-0.4,0) {};
    \coordinate (h2) at (0.4,0) {};
    \coordinate (h3) at (1.6,0) {};
    \coordinate (h4) at (2.4,0) {};
    \coordinate (h5) at (3.6,0) {};
    \coordinate (h6) at (4.4,0) {};
    \coordinate (h7) at (5,0) {};
    \coordinate (u0) at (0,0.4) {};
    \coordinate (u1) at (0,1.5) {};
    \coordinate (u2) at (4,0.4) {};
    \coordinate (u3) at (4,1.5) {};
    \coordinate (d0) at (2,-0.4) {};
    \coordinate (d1) at (2,-1.1) {};
    \coordinate (d2) at (2,-1.9) {};
    \coordinate (d3) at (2,-2.3) {};
    \draw[color=black] (h0) -- (h1);
    \draw[color=black] (h2) -- (h3);
    \draw[color=black] (h4) -- (h5);
    \draw[color=black] (h6) -- (h7);
    \draw[color=black] (u0) -- (u1);
    \draw[color=black] (u2) -- (u3);
    \draw[color=black] (d0) -- (d1);
    \draw[color=black] (d2) -- (d3);
\end{scope}
\end{tikzpicture}} \quad = \quad
\raisebox{-.5\height}{
\begin{tikzpicture}[scale=1.0]
\begin{scope}[shift={({0},{0})}]
 \node (node1) at (0,0) [thick,minimum size=.15cm, minimum width = 0.8cm,circle,draw] {$A$};
 \node (node2) at (2,0) [thick,minimum size=.15cm, minimum width = 0.8cm,circle,draw] {$B$};
 \node (node3) at (4,0) [thick,minimum size=.15cm, minimum width = 0.8cm,circle,draw] {$A$};
 \node (node4) at (0,1.5) [thick,minimum size=.15cm, minimum width = 0.8cm,circle,draw] {$X$};
 \node (node5) at (4,1.5) [thick,minimum size=.15cm, minimum width = 0.8cm,circle,draw] {$X$};
    \coordinate (h0) at (-1,0) {};
    \coordinate (h1) at (-0.4,0) {};
    \coordinate (h2) at (0.4,0) {};
    \coordinate (h3) at (1.6,0) {};
    \coordinate (h4) at (2.4,0) {};
    \coordinate (h5) at (3.6,0) {};
    \coordinate (h6) at (4.4,0) {};
    \coordinate (h7) at (5,0) {};
    \coordinate (u0) at (0,0.4) {};
    \coordinate (u1) at (0,1.1) {};
    \coordinate (u2) at (4,0.4) {};
    \coordinate (u3) at (4,1.1) {};
    \coordinate (u4) at (0,1.9) {};
    \coordinate (u5) at (0,2.3) {};
    \coordinate (u6) at (4,1.9) {};
    \coordinate (u7) at (4,2.3) {};
    \coordinate (d0) at (2,-0.4) {};
    \coordinate (d1) at (2,-1.5) {};
    \draw[color=black] (h0) -- (h1);
    \draw[color=black] (h2) -- (h3);
    \draw[color=black] (h4) -- (h5);
    \draw[color=black] (h6) -- (h7);
    \draw[color=black] (u0) -- (u1);
    \draw[color=black] (u2) -- (u3);
    \draw[color=black] (d0) -- (d1);
    \draw[color=black] (u4) -- (u5);
    \draw[color=black] (u6) -- (u7);
\end{scope}
\end{tikzpicture}} \ .
\end{align}

Applying the KW MPO to the XXZ $\check R$-matrix gives
\begin{align}
    \mathcal{D}_{\rm KW} \check{\mathbf{R}}_{j,j+1} (u) = \check{\mathtt{R}}_{j-1/2 , j+1/2 ,j+3/2}(u) \mathcal{D}_{\rm KW} \; ,
\label{eq:Rcheckduality}
\end{align}
where the transformed $\check{R}$-matrix remains local and acts on three neighbouring sites. 
The left-hand and right-hand sides are shown diagrammatically as
\begin{align}
\raisebox{-.5\height}{
\begin{tikzpicture}[scale=1.0]
\begin{scope}[shift={({0},{0})}]
 \node (node1) at (0,0) [thick,minimum size=.15cm, minimum width = 0.8cm,circle,draw] {$A$};
 \node (node2) at (2,0) [thick,minimum size=.15cm, minimum width = 0.8cm,circle,draw] {$B$};
 \node (node3) at (4,0) [thick,minimum size=.15cm, minimum width = 0.8cm,circle,draw] {$A$};
  \node (node4) at (6,0) [thick,minimum size=.15cm, minimum width = 0.8cm,circle,draw] {$B$};
  \node (node5) at (8,0) [thick,minimum size=.15cm, minimum width = 0.8cm,circle,draw] {$A$};
  \node (node6) at (4,-2) {$\check{\mathbf{R}}$};
    \coordinate (h0) at (-1,0) {};
    \coordinate (h1) at (-0.4,0) {};
    \coordinate (h2) at (0.4,0) {};
    \coordinate (h3) at (1.6,0) {};
    \coordinate (h4) at (2.4,0) {};
    \coordinate (h5) at (3.6,0) {};
    \coordinate (h6) at (4.4,0) {};
    \coordinate (h7) at (5.6,0) {};
    \coordinate (h8) at (6.4,0) {};
    \coordinate (h9) at (7.6,0) {};
    \coordinate (h10) at (8.4,0) {};
    \coordinate (h11) at (9,0) {};
    \coordinate (u0) at (0,0.4) {};
    \coordinate (u1) at (0,1) {};
    \coordinate (u2) at (4,0.4) {};
    \coordinate (u3) at (4,1) {};
    \coordinate (u4) at (8,0.4) {};
    \coordinate (u5) at (8,1) {};
    \coordinate (d0) at (2,-0.4) {};
    \coordinate (d1) at (2,-1) {};
    \coordinate (d2) at (6,-0.4) {};
    \coordinate (d3) at (6,-1) {};
    \coordinate (m1) at (3.5,-1.5) {};
    \coordinate (m2) at (4.5,-1.5) {};
    \coordinate (m3) at (3.5,-2.5) {};
    \coordinate (m4) at (4.5,-2.5) {};
    \coordinate (m5) at (3,-3) {};
    \coordinate (m6) at (5,-3) {};
    \coordinate (r1) at (4,-1) {};
    \coordinate (r2) at (3,-2) {};
    \coordinate (r3) at (4,-3) {};
    \coordinate (r4) at (5,-2) {};
    \draw[color=black] (h0) -- (h1);
    \draw[color=black] (h2) -- (h3);
    \draw[color=black] (h4) -- (h5);
    \draw[color=black] (h6) -- (h7);
    \draw[color=black] (h8) -- (h9);
    \draw[color=black] (h10) -- (h11);
    \draw[color=black] (u0) -- (u1);
    \draw[color=black] (u2) -- (u3);
    \draw[color=black] (u4) -- (u5);
    \draw[color=black] (d0) -- (d1);
    \draw[color=black] (d2) -- (d3);
    \draw[color=black] (r1) -- (r2);
    \draw[color=black] (r2) -- (r3);
    \draw[color=black] (r3) -- (r4);
    \draw[color=black] (r4) -- (r1);
    \draw[color=black] (d1) -- (m1);
    \draw[color=black] (d3) -- (m2);
    \draw[color=black] (m3) -- (m5);
    \draw[color=black] (m4) -- (m6);
\end{scope}
\end{tikzpicture}}\ ,
\end{align}

\begin{align}
\raisebox{-.5\height}{
\begin{tikzpicture}[scale=1.0]
\begin{scope}[shift={({0},{0})}]
 \node (node1) at (0,0) [thick,minimum size=.15cm, minimum width = 0.8cm,circle,draw] {$A$};
 \node (node2) at (2,0) [thick,minimum size=.15cm, minimum width = 0.8cm,circle,draw] {$B$};
 \node (node3) at (4,0) [thick,minimum size=.15cm, minimum width = 0.8cm,circle,draw] {$A$};
  \node (node4) at (6,0) [thick,minimum size=.15cm, minimum width = 0.8cm,circle,draw] {$B$};
  \node (node5) at (8,0) [thick,minimum size=.15cm, minimum width = 0.8cm,circle,draw] {$A$};
  \node (node6) at (4,2) {$\check{\mathtt{R}}$};
    \coordinate (h0) at (-1,0) {};
    \coordinate (h1) at (-0.4,0) {};
    \coordinate (h2) at (0.4,0) {};
    \coordinate (h3) at (1.6,0) {};
    \coordinate (h4) at (2.4,0) {};
    \coordinate (h5) at (3.6,0) {};
    \coordinate (h6) at (4.4,0) {};
    \coordinate (h7) at (5.6,0) {};
    \coordinate (h8) at (6.4,0) {};
    \coordinate (h9) at (7.6,0) {};
    \coordinate (h10) at (8.4,0) {};
    \coordinate (h11) at (9,0) {};
    \coordinate (u0) at (0,0.4) {};
    \coordinate (u1) at (0,3.5) {};
    \coordinate (u2) at (4,0.4) {};
    \coordinate (u3) at (4,1) {};
    \coordinate (u4) at (8,0.4) {};
    \coordinate (u5) at (8,3.5) {};
    \coordinate (d0) at (2,-0.4) {};
    \coordinate (d1) at (2,-1) {};
    \coordinate (d2) at (6,-0.4) {};
    \coordinate (d3) at (6,-1) {};
    \coordinate (g1) at (4,1) {};
    \coordinate (g2) at (0,2) {};
    \coordinate (g3) at (4,3) {};
    \coordinate (g4) at (8,2) {};
    \coordinate (g5) at (4,3.5) {};
    \draw[color=black] (h0) -- (h1);
    \draw[color=black] (h2) -- (h3);
    \draw[color=black] (h4) -- (h5);
    \draw[color=black] (h6) -- (h7);
    \draw[color=black] (h8) -- (h9);
    \draw[color=black] (h10) -- (h11);
    \draw[color=black] (u0) -- (u1);
    \draw[color=black] (u2) -- (u3);
    \draw[color=black] (u4) -- (u5);
    \draw[color=black] (d0) -- (d1);
    \draw[color=black] (d2) -- (d3);
    \draw[color=black] (g1) -- (g2);
    \draw[color=black] (g2) -- (g3);
    \draw[color=black] (g3) -- (g4);
    \draw[color=black] (g4) -- (g1);
    \draw[color=black] (g3) -- (g5);
\end{scope}
\end{tikzpicture}}\ ,
\end{align}
respectively.

Explicitly, the dual $\check{R}$-matrix reads 
\begin{align}
    \check{\mathtt{R}}_{j+\frac{1}{2}}(u) : = \check{\mathtt{R}}_{j-\frac{1}{2} , j+\frac{1}{2} ,j+\frac{3}{2}}(u) = \sin \mu \, \mathbb{I} + (\sin(u+\mu) - \sin \mu) \tilde{\mathbf{P}}_{j+\frac{1}{2}} + \sin u \,  \tilde{\mathbf{S}}_{j+\frac{1}{2}}  \;,
\end{align}
where the operators $\tilde{\mathbf{S}}_{j+\frac{1}{2}}$ and $\tilde{\mathbf{P}}_{j+\frac{1}{2}}$ are defined as
\begin{align}
    \tilde{\mathbf{S}}_{j+\frac{1}{2}} = \frac{1}{2} Z_{j+\frac{1}{2}} \left( \mathbb{I} - X_{j-\frac{1}{2}}X_{j+\frac{3}{2}} \right) \; , \quad \tilde{\mathbf{P}}_{j+\frac{1}{2}} = \frac{1}{2} \left( \mathbb{I} + X_{j-\frac{1}{2}} X_{j+\frac{3}{2}} \right) \; .
\end{align}
Notice that both $\tilde{\mathbf{S}}$ and $\tilde{\mathbf{P}}$, and thus  $\check{\mathtt{R}}$, act diagonally on the first and third index when written in the $X$ basis, i.e.,
\begin{align}
    \check{\mathtt{R}} (u) =  \sum_{j,m,n,k = \pm} \check{\mathtt{R}}_{j,m,k}^{j,n,k} (u) \ket{j,n,k} \bra{j,m,k} .
\end{align}
The operators $\tilde{\mathbf{S}}$ and $\tilde{\mathbf{P}}$ also form a representation of the same chromatic algebra as the XXZ model (but this representation is different from the original one). Therefore the dual $\check{R}$-matrix satisfies the Yang--Baxter equation~\eqref{eq:check_R_YBE},
\begin{align}
    \check{\mathtt{R}}_{j+\frac{1}{2}} (u-v) \check{\mathtt{R}}_{j+\frac{3}{2}} (u) \check{\mathtt{R}}_{j+\frac{1}{2}} (v) = \check{\mathtt{R}}_{j+\frac{3}{2}} (v) \check{\mathtt{R}}_{j+\frac{1}{2}} (u) \check{\mathtt{R}}_{j+\frac{3}{2}} (u-v) \; .
\end{align}
Thus the local integrable structure survives the non-invertible duality transformation. In particular, both Hamiltonians arise from derivatives of their corresponding $\check R$-matrices,
\begin{align}
    \mathbf{H}_{\rm XXZ} = \sum_{j=1}^\ell 2 \, \check{\mathbf{R}}_{j,j+1}' (0) \; , \quad \mathbf{H}_{\rm Izz} = \sum_{j=1}^\ell 2 \, \check{\mathtt{R}}_{j+1/2}' (0) \; .
\end{align}

Having established the Yang--Baxter structure of the dual model, we now construct the corresponding commuting transfer matrices. Using the 3-body transformed $\check{R}$-matrix, we can define a 2-body operator, the associated face-type $R$-matrix,
\begin{align}
\bra{j,n} \mathtt{R}(u) \ket{m,k} 
    = \check{\mathtt{R}}_{j,m,k}^{j,n,k} (u) := \mathtt{R}_{j,n}^{m,k} (u)\; .
\label{eq:IzzRmat}
\end{align}
Graphically, this corresponds to rotating the transformed $\check{R}$-matrix by $\pi/4$, 
\begin{align}
\bra{j,n} \mathtt{R}(u) \ket{m,k} = 
\raisebox{-.5\height}{
\begin{tikzpicture}[scale=1.0]
\begin{scope}[shift={({0},{0})}]
  \node (node1) at (1,1) {$\mathtt{R}(u)$};
  \node (node2) at (0,-0.3) {$m$};
  \node (node3) at (0,2.3) {$j$};
  \node (node2) at (2,-0.3) {$k$};
  \node (node3) at (2,2.3) {$n$};
    \coordinate (a) at (0,0) {};
    \coordinate (b) at (2,0) {};
    \coordinate (c) at (0,2) {};
    \coordinate (d) at (2,2) {};
    \draw[color=black] (a) -- (b);
    \draw[color=black] (a) -- (c);
    \draw[color=black] (b) -- (d);
    \draw[color=black] (c) -- (d);
\end{scope}
\end{tikzpicture}} .
\end{align}
We remind the reader that the basis of local vector spaces is in the eigenbasis of the $X$ operator $\{ |+ \rangle , |- \rangle\}$.

The transfer matrix is constructed as a \emph{split-index matrix product operator} \cite{Stephen25} using the $R$-matrix \eqref{eq:IzzRmat}, 
\begin{align}
 T_{\rm Izz} = \sum_{i,j} \mathtt{R}_{i_1,i_2}^{j_1,j_2}(u) \cdot \mathtt{R}_{i_2,i_3}^{j_2,j_3}(u) \dots \mathtt{R}_{i_\ell,i_1}^{j_\ell,j_1}(u) \ket{j_1 j_2 \dots j_\ell} \bra{i_1,i_2, \dots, i_\ell}\ ,
\end{align}
which can be depicted as 
\begin{align}
\bra{j_1, j_2, \dots, j_\ell} T_{\rm Izz} (u) \ket{i_1, i_2, \dots, i_\ell} = \quad 
\raisebox{-.5\height}{
\begin{tikzpicture}[scale=1.0]
\begin{scope}[shift={({0},{0})}]
  \node (node1) at (0,-0.3) {$i_1$};
  \node (node2) at (2,-0.3) {$i_2$};
  \node (node3) at (3,-0.3) {$\cdots$};
  \node (node4) at (0,2.3) {$j_1$};
  \node (node5) at (2,2.3) {$j_2$};
  \node (node6) at (3,2.3) {$\cdots$};
  \node (node7) at (1,1) {$\mathtt{R}(u)$};
  \node (node8) at (3,1) {$\mathtt{R}(u)$};
  \node (node9) at (5,1) {$\mathtt{R}(u)$};
    \coordinate (a1) at (-1,0) {};
    \coordinate (a2) at (0,0) {};
    \coordinate (a3) at (2,0) {};
    \coordinate (a4) at (4,0) {};
    \coordinate (a5) at (6,0) {};
    \coordinate (a6) at (7,0) {};
    \coordinate (b1) at (-1,2) {};
    \coordinate (b2) at (0,2) {};
    \coordinate (b3) at (2,2) {};
    \coordinate (b4) at (4,2) {};
    \coordinate (b5) at (6,2) {};
    \coordinate (b6) at (7,2) {};
    \draw[color=black] (a1) -- (a6);
    \draw[color=black] (b1) -- (b6);
    \draw[color=black] (a2) -- (b2);
    \draw[color=black] (a3) -- (b3);
    \draw[color=black] (a4) -- (b4);
    \draw[color=black] (a5) -- (b5);
\end{scope}
\end{tikzpicture}} \; .
\label{eq:Izztransfer}
\end{align}

In terms of the language of integrability, the transfer matrix $T_{\rm Izz}$ is a transfer matrix of a face-type model, compared to the transfer matrix of the vertex-type $T_{\rm XXZ}$. Therefore, the duality transformation $\mathcal{D}_{\rm KW}$ can be seen as a vertex-face correspondence, similar to the one of Baxter between the 6-vertex model and the SOS model~\cite{Baxter:1972wh, Kojima:2005qz, Ikhlef:2016sgm}.

The transfer matrix \eqref{eq:Izztransfer} is \emph{regular}, i.e., $T_{\rm Izz} (0) $ is the translation:
\begin{equation}
\begin{split}
T_{\rm Izz} (0) & \propto \sum_{i_1,i_2, \dots i_{\ell}} | i_{\ell}, i_1, \dots, i_{\ell-1} \rangle \langle i_1 , i_2 \dots , i_{\ell} | \\ 
& = \sum_{i_1,i_2, \dots i_{\ell}}  \quad 
\raisebox{-.5\height}{
\begin{tikzpicture}[scale=1.0]
\begin{scope}[shift={({0},{0})}]
  \node (node1) at (0,-0.3) {$i_1$};
  \node (node2) at (2,-0.3) {$i_2$};
  \node (node3) at (3,-0.3) {$\cdots$};
  \node (node4) at (0,2.3) {$i_{\ell}$};
  \node (node5) at (2,2.3) {$i_1$};
  \node (node6) at (3,2.3) {$\cdots$};
  \node (node7) at (6,-0.3) {$i_\ell$};
  \node (node8) at (6,2.3) {$i_{\ell-1}$};
    \coordinate (a1) at (0,0) {};
    \coordinate (a2) at (2,2) {};
    \coordinate (a3) at (2,0) {};
    \coordinate (a4) at (4,2) {};
    \coordinate (a5) at (4,0) {};
    \coordinate (a6) at (6,2) {};
    \coordinate (a7) at (6,0) {};
    \coordinate (a8) at (6.5,0.5) {};
    \coordinate (a0) at (-0.5,1.5) {};
    \coordinate (a9) at (0,2) {};
    \draw[color=black] (a1) -- (a2);
    \draw[color=black] (a3) -- (a4);
    \draw[color=black] (a5) -- (a6);
    \draw[color=black] (a7) -- (a8);
    \draw[color=black] (a0) -- (a9);
\end{scope}
\end{tikzpicture}} = \prod_{n=1}^{\ell-1} \mathbf{SW}_{n,n+1} \; .
\end{split}
\end{equation} (Recall the definition of \textbf{SW} in \cref{eq:SW}.)
In addition, the $R$-matrix satisfies the Yang--Baxter equation,
\begin{align}
\sum_i
\raisebox{-.5\height}{
\begin{tikzpicture}[scale=1.0]
\begin{scope}[shift={({0},{0})}]
  \node (node1) at (0,0) {$\mathtt{R}(u-v)$};
  \node (node2) at (2,1) {$\mathtt{R}(u)$};
  \node (node3) at (2,-1) {$\mathtt{R}(v)$};
  \node (node4) at (1.5,0) {$i$};
    \coordinate (a) at (0,2) {};
    \coordinate (b) at (2,0) {};
    \coordinate (c) at (0,-2) {};
    \coordinate (d) at (-2,0) {};
    \coordinate (e) at (2,2) {};
    \coordinate (f) at (4,0) {};
    \coordinate (g) at (2,-2) {};
    \draw[color=black] (a) -- (b);
    \draw[color=black] (b) -- (c);
    \draw[color=black] (c) -- (d);
    \draw[color=black] (a) -- (d);
    \draw[color=black] (a) -- (e);
    \draw[color=black] (e) -- (f);
    \draw[color=black] (f) -- (g);
    \draw[color=black] (b) -- (f);
    \draw[color=black] (c) -- (g);
\end{scope}
\end{tikzpicture}} = \sum_j 
\raisebox{-.5\height}{
\begin{tikzpicture}[scale=1.0]
\begin{scope}[shift={({0},{0})}]
  \node (node1) at (0,0) {$\mathtt{R}(u-v)$};
  \node (node2) at (-2,1) {$\mathtt{R}(v)$};
  \node (node3) at (-2,-1) {$\mathtt{R}(u)$};
  \node (node4) at (-1.5,0) {$j$};
    \coordinate (a) at (0,2) {};
    \coordinate (b) at (-2,0) {};
    \coordinate (c) at (0,-2) {};
    \coordinate (d) at (2,0) {};
    \coordinate (e) at (-2,2) {};
    \coordinate (f) at (-4,0) {};
    \coordinate (g) at (-2,-2) {};
    \draw[color=black] (a) -- (b);
    \draw[color=black] (b) -- (c);
    \draw[color=black] (c) -- (d);
    \draw[color=black] (a) -- (d);
    \draw[color=black] (a) -- (e);
    \draw[color=black] (e) -- (f);
    \draw[color=black] (f) -- (g);
    \draw[color=black] (b) -- (f);
    \draw[color=black] (c) -- (g);
\end{scope}
\end{tikzpicture}}\; .
\end{align}
Therefore, the transfer matrices of the dual model commute with each other using the same train argument as in the vertex model case (see \cref{sec:commuting}):
\begin{align}
    \left[ T_{\rm Izz} (u) , T_{\rm Izz} (v) \right] = 0 \; , \quad \forall ~ u,v \in \mathbb{C} \; .
\end{align}

We justify the integrability of the Hamiltonian \eqref{eq:IzzHamiltonian} by taking the logarithmic derivatives of the transfer matrix \eqref{eq:Izztransfer}, which is obtained in the same way as the vertex model case (i.e. the XXZ Hamiltonian),
\begin{align}
    \mathbf{H}_{\rm Izz} = \left. 2\sin \mu \, \partial_u T_{\rm Izz} (u) \right|_{u=0} = \sum_{j=1}^{\ell} 2\, \check{\mathtt{R}}_{j+\frac{1}{2}}^\prime (0) \; ,
\end{align}
since the Hamiltonian commutes with the transfer matrix,
\begin{align}
    \left[ \mathbf{H}_{\rm Izz} , T_{\rm Izz} (u) \right] = 0 \; , \quad \forall u \in \mathbb{C} \; .
\end{align}

The KW duality operator not only intertwines the two $\check{R}$-matrices of the XXZ model and its dual, but also their transfer matrices, 
\begin{align}
    \mathcal{D}_{\rm KW} T_{\rm XXZ} (u) = T_{\rm Izz} (u) \mathcal{D}_{\rm KW} \; .
\label{eq:transfermatduality}
\end{align}

Unlike the duality transformation of the $\check{R}$-matrices \eqref{eq:Rcheckduality} that can be proven directly using the action of the duality operator, we are only able to prove the duality transformation of the transfer matrices \eqref{eq:transfermatduality} by expanding the logarithm of the transfer matrices order by order, i.e.
\begin{align}
    \log T_{\rm XXZ} (u) = \sum_{n=0}^{\infty} \frac{u^n}{n !} \mathbf{Q}_n , \quad \log T_{\rm Izz} (u) = \sum_{n=0}^{\infty} \frac{u^n}{n !} \mathtt{Q}_n \; ,
\end{align}
where 
\begin{align}
    \mathbf{Q}_n = \sum_j f_n(\check{\mathbf{R}}_{j,j+1}) , \quad \mathtt{Q}_n = \sum_j f_n(\check{\mathtt{R}}_{j+\frac{1}{2}}) \; .
\end{align}

It is straightforward to verify that
\begin{align}
    \mathcal{D}_{\rm KW} \mathbf{Q}_n = \mathtt{Q}_n \mathcal{D}_{\rm KW} \; , \quad \forall n \in \mathbb{Z}_{\geq 0} \; ,
\end{align}
using \eqref{eq:Rcheckduality}. Therefore Eq.~\eqref{eq:transfermatduality} follows order by order in the spectral parameter.

We emphasize that the construction of the transfer matrix $T_{\rm Izz}(u)$ for the Ising zig-zag model, together with the intertwining relation~\eqref{eq:transfermatduality} appears to be new. In particular, while the Kramers--Wannier duality and vertex-face correspondences are well established in integrable lattice models \cite{Baxter:1972wh, Kojima:2005qz, Ikhlef:2016sgm}, the explicit realization of the dual Ising zig-zag transfer matrix as a local MPO obtained from the dual $\check R$-matrix, together with its commuting family derived through baxterization, has not previously been formulated in this manner. The relation above therefore provides a direct operator-level correspondence between the XXZ transfer matrix and the transfer matrix of the dual face-type model.

\section{MPO transformed lattice models---general observations}
\label{sec:conclusion}

In this section we discuss how we can apply ideas found in the case studies more generally. In particular, we explain how the modified YBA that we saw in the XXZ case applies to any Yang--Baxter integrable model transformed by an MPO with exact MPO inverse. We then make some general remarks about the connection to charge pumps and the case of non-invertible dualities.

In order to do this, let us first collect some useful results about MPOs and MPUs.

\subsection{Properties of MPO that have an exact MPO inverse}\label{sec:MPO}
Let us consider invertible MPOs where the inverse is another MPO with finite bond dimension\footnote{This is a restriction: there exist examples of invertible MPO where the inverse is not an MPO \cite{Ilievski15,Ilievski16}. One way to see this is that the MPO transfer matrix in the references leads to quasilocal charges, a contradiction to the result of \cref{sec:locality}.}. This class includes all MPUs as a subfamily.

We restrict to translation-invariant MPO, defined by a single tensor $A\in \mathcal{M}_D \otimes \mathcal{M}_d$ ($\mathcal{M}_D$ denotes the vector space of $D\times D$ matrices), 
\begin{align}
    A = \sum_{i,j=1}^d A_{ij} \otimes \ket{i}\bra{j}\ .
\end{align}
The MPO on $n$ sites is an operator $O_n(A)\in \mathcal{M}_d^{\otimes n}$ defined as 
\begin{align}
  O_n(A) = \sum_{\mathbf{i},\mathbf{j}} \tr (A_{i_1,j_1} A_{i_2,j_2} \dots A_{i_n,j_n}) \ket{i_1 i_2\dots i_n}\bra{j_1 j_2 \dots j_n}.
\end{align}
Graphically, this operator is represented as 
\begin{align}\label{eq:ABMPO}
    O_n(A) = 
    \begin{tikzpicture}[baseline = -1mm, yscale=0.6]
        \draw (-0.7,0) rectangle (3.7, -1.2);
        \foreach \x in {0,1,3} {
            \draw (\x,-1) --++ (0,2);
            \node[tensor] at (\x,0) {$A$};
        }
        \node[fill=white] at (2,0) {$\dots$};
        \draw[decoration={brace, raise=1mm,amplitude=1mm}, decorate] (-0.1, 1) -- (3.1,1) node[midway, anchor=south, yshift=1.5mm] {$n$}; 
    \end{tikzpicture} \ .
\end{align}

Note that considering MPOs that are invertible and whose inverse is also an MPO means that we assume that for the MPO tensor $A\in \mathcal{M}_D \otimes \mathcal{M}_d$ there is another MPO tensor $B\in \mathcal{M}_{D'} \otimes \mathcal{M}_d$ such that $O_n(A) \cdot O_n(B) = O_n(B)\cdot O_n(A) = \mathbb{I}^{\otimes n}$, that is,
\begin{equation}
    O_n(A)^{-1} = O_n(B) = 
    \begin{tikzpicture}[baseline = -1mm, yscale=0.6]
        \draw (-0.7,0) rectangle (3.7, -1.2);
        \foreach \x in {0,1,3} {
            \draw (\x,-1) --++ (0,2);
            \node[tensor] at (\x,0) {$B$};
        }
        \node[fill=white] at (2,0) {$\dots$};
        \draw[decoration={brace, raise=1mm,amplitude=1mm}, decorate] (-0.1, 1) -- (3.1,1) node[midway, anchor=south, yshift=1.5mm] {$n$}; 
    \end{tikzpicture} \ .
\end{equation}

Given this set up we have the following four results, proved in \cref{app:MPO}. While the proofs follow relatively straightforwardly from existing literature \cite{Molnar18}, we expect these results to be of independent interest.
\begin{shaded}
\begin{result}[$A$-$B$ transfer matrix decomposition]\label{sec:ABtransfer}
We obtain that the $A$-$B$ `transfer matrix', defined as the contraction of the two different MPO tensors, has the following decomposition:
\begin{align}\label{eq:mpo_transfer}
    \begin{tikzpicture}[baseline=4mm,background rectangle/.style=
{fill=white},
   show background rectangle]
        \draw[shorten <>=2mm] (0,-1) -- (0,2);
        \draw[shorten <>=2mm] (-1,0) -- (1,0);
        \draw[shorten <>=2mm] (-1,1) -- (1,1);
        \node[tensor] at (0,0) {$B$};
        \node[tensor] at (0,1) {$A$};
        \node[] at (1.5,0.5) {$=$};
        \begin{scope}[shift={(4,0)}]
        \draw[shorten <>=2mm] (0,-1) -- (0,2);
        \draw[shorten <>=2mm] (-2,1) -- (-1,1) -- (-1,0) -- (-2,0);
        \draw[shorten <>=2mm] (2,1) -- (1,1) -- (1,0) -- (2,0);
        \node[tensor] at (-1,0.5) {$v_R$};
        \node[tensor] at (1,0.5) {$v_L$};
        \node[] at (2.5,0.5) {$+$};
        \begin{scope}[shift={(4,0)}]
        \draw[shorten <>=2mm] (0,-1) -- (0,2);
        \draw[shorten <>=2mm] (-1,0) -- (1,0);
        \draw[shorten <>=2mm] (-1,1) -- (1,1);
        \node[tensor, minimum height=1.5cm] at (0,0.5) {$N$};\end{scope}\end{scope}
    \end{tikzpicture} \ , 
\end{align}
where for any system size $k$,
\begin{align}
    \begin{tikzpicture}[baseline = 4mm,background rectangle/.style=
{fill=white},
   show background rectangle]
        \draw (-1,1) rectangle (4, 0);
        \foreach \x in {0,1,3} {
            \draw (\x,-0.7) -- (\x,1.7);
            \node[tensor, minimum height=1.5cm] at (\x,0.5) {$N$};
        }
        \node[fill=white] at (2,1) {$\dots$};
        \node[fill=white] at (2,0) {$\dots$};
        \node[tensor] at (-1,0.5) {$v_L$};
        \node[tensor] at (4,0.5) {$v_R$};
        \draw[decoration={brace, mirror,raise=1mm,amplitude=1mm}, decorate] (-0.1,-0.7) -- (3.1,-0.7) node[midway, anchor=north, yshift=-1.5mm] {$k$}; 
        \node[] at (5,0.5) {$=0$};
    \end{tikzpicture}\ , \label{eq:n-orthogonal}
\end{align}
and $N$ is \emph{nilpotent}, i.e.,  there is $n_0\in \mathbb{N}$ such that for all $k>n_0$, the following contraction of $k$ MPO tensors is zero:
\begin{align}
        \begin{tikzpicture}[baseline=4mm,background rectangle/.style=
{fill=white},
   show background rectangle]
        \draw[shorten <>=2mm] (0,-1) -- (0,2);
        \draw[shorten <>=2mm] (2,-1) -- (2,2);
        \draw[shorten <>=2mm] (-1,0) -- (3,0);
        \draw[shorten <>=2mm] (-1,1) -- (3,1);
        \node[tensor, minimum height=1.5cm] at (0,0.5) {$N$};
        \node[tensor, minimum height=1.5cm] at (2,0.5) {$N$};
        \node[fill=white] at (1,0) {$\dots$};
        \node[fill=white] at (1,1) {$\dots$};
        \draw[decoration={brace, mirror,raise=1mm,amplitude=1mm}, decorate] (-0.1,-0.8) -- (2.1,-0.8) node[midway, anchor=north, yshift=-1.5mm] {$k$}; 
        \node[] at (4,0.5) {$=0$};
    \end{tikzpicture}  \ . \label{eq:nilpotency}
\end{align}
\end{result}\end{shaded}
Note that \cref{eq:mpo_transfer} is the analogue of \cref{eq:MPUdecomposition} in the MPU case \cite{Cirac17,Sahinoglu18}.
A consequence of these properties is that for any system size $k$, 
\begin{align}
    \begin{tikzpicture}[baseline = 4mm]
        \draw (-1,1) rectangle (4, 0);
        \foreach \x in {0,1,3} {
            \draw (\x,-0.7) -- (\x,1.7);
            \node[tensor] at (\x,0) {$B$};
            \node[tensor] at (\x,1) {$A$};
        }
        \node[fill=white] at (2,1) {$\dots$};
        \node[fill=white] at (2,0) {$\dots$};
        \node[tensor] at (-1,0.5) {$v_L$};
        \node[tensor] at (4,0.5) {$v_R$};
        \draw[decoration={brace, mirror,raise=1mm,amplitude=1mm}, decorate] (-0.1,-0.7) -- (3.1,-0.7) node[midway, anchor=north, yshift=-1.5mm] {$k$}; 
    \end{tikzpicture} \ = \mathbb{I}^{\otimes k}\ ,
\end{align}
and that the matrix $v_R$ ($v_L$) can be pushed one position to the right (left) as long as there are at least $k+1 > n_0$  transfer matrices on its right (left):
\begin{align}\label{eq:MPOpullthrough}
    \begin{tikzpicture}[baseline = 4mm]
        \draw (4,1 ) -- (-1,1) -- (-1,0) -- (4, 0);
        \foreach \x in {0,1,3} {
            \draw (\x,-0.7) -- (\x,1.7);
            \node[tensor] at (\x,0) {$B$};
            \node[tensor] at (\x,1) {$A$};
        }
        \node[fill=white] at (2,1) {$\dots$};
        \node[fill=white] at (2,0) {$\dots$};
        \node[tensor] at (-1,0.5) {$v_L$};
        \draw[decoration={brace, mirror,raise=1mm,amplitude=1mm}, decorate] (-0.1,-0.7) -- (3.1,-0.7) node[midway, anchor=north, yshift=-1.5mm] {$k+1$}; 
    \end{tikzpicture} \ = \mathbb{I} \otimes 
    \begin{tikzpicture}[baseline = 4mm]
        \draw (4,1 ) -- (0,1) -- (0,0) -- (4, 0);
        \foreach \x in {1,3} {
            \draw (\x,-0.7) -- (\x,1.7);
            \node[tensor] at (\x,0) {$B$};
            \node[tensor] at (\x,1) {$A$};
        }
        \node[fill=white] at (2,1) {$\dots$};
        \node[fill=white] at (2,0) {$\dots$};
        \node[tensor] at (0,0.5) {$v_L$};
        \draw[decoration={brace, mirror,raise=1mm,amplitude=1mm}, decorate] (1-0.1,-0.7) -- (3.1,-0.7) node[midway, anchor=north, yshift=-1.5mm] {$k$}; 
    \end{tikzpicture} \ , \\
    \begin{tikzpicture}[baseline = 4mm, xscale=-1]
        \draw (4,1 ) -- (-1,1) -- (-1,0) -- (4, 0);
        \foreach \x in {0,1,3} {
            \draw (\x,-0.7) -- (\x,1.7);
            \node[tensor] at (\x,0) {$B$};
            \node[tensor] at (\x,1) {$A$};
        }
        \node[fill=white] at (2,1) {$\dots$};
        \node[fill=white] at (2,0) {$\dots$};
        \node[tensor] at (-1,0.5) {$v_R$};
        \draw[decoration={brace, raise=1mm,amplitude=1mm}, decorate] (-0.1,-0.8) -- (3.1,-0.8) node[midway, anchor=north, yshift=-1.5mm] {$k+1$}; 
    \end{tikzpicture} \ =  
    \begin{tikzpicture}[baseline = 4mm, xscale=-1]
        \draw (4,1 ) -- (0,1) -- (0,0) -- (4, 0);
        \foreach \x in {1,3} {
            \draw (\x,-0.7) -- (\x,1.7);
            \node[tensor] at (\x,0) {$B$};
            \node[tensor] at (\x,1) {$A$};
        }
        \node[fill=white] at (2,1) {$\dots$};
        \node[fill=white] at (2,0) {$\dots$};
        \node[tensor] at (0,0.5) {$v_R$};
        \draw[decoration={brace, raise=1mm,amplitude=1mm}, decorate] (1-0.1,-0.8) -- (3.1,-0.8) node[midway, anchor=north, yshift=-1.5mm] {$k$}; 
    \end{tikzpicture}  \otimes \mathbb{I}\ .
\end{align}
This gives the pulling through equation \eqref{eq:pullingthrough} for such MPO if we block the physical sites.

Our next statement generalizes to our setting the result that MPU are equivalent to quantum cellular automata \cite{Cirac17,Sahinoglu18}.

\begin{shaded}
\begin{result}[Locality preservation]\label{sec:locality}
Consider now any local operator $X$, supported on $k$ sites, and conjugate by the MPO $O_n(A)$ (where $n$ is sufficiently large). We have that
\begin{align}
    O_n(A) X O_n(A)^{-1} = O_n(A) X O_n(B)
\end{align}
is a local operator, and the size of its support is $k+2n_0$ ($n_0$ is the nilpotency length \eqref{eq:nilpotency}).  \end{result}\end{shaded}

For the next result,  we further assume that the MPO tensor depends (differentiably) on a parameter $t$: $A=A_t$, and for all $t$ there is an inverse MPO described by the tensor $B_t$. Then we have that the logarithmic derivative of an MPO with MPO inverse is local.  More precisely we have:
\begin{shaded}
\begin{result}[Generator of local charges]
Given the above assumptions, we have that the operator
\begin{align}
    \partial_t O_n(A_t) \cdot O_n(B_t)
\end{align}
is a sum of terms, where each term is at most $(2n_0+1)$-local. Here $n_0$ is the nilpotency length associated with the $A$-$B$ transfer matrix at the fixed time $t$.\end{result}\end{shaded}
Finally, we give a useful lemma that shows that there are no non-trivial MPO with MPO inverse where the extended $R$-matrix automatically solves the (unmodified) RLL relation.\begin{shaded}
\begin{result}[MPO triviality when $N$ annihilates the dominant left-right eigenvectors]\label{sec:MPUtrivial}
Suppose that the decomposition in \cref{sec:ABtransfer} satisfies
\begin{align}\label{eq:mpu_simple_assumption}
    \begin{tikzpicture}[baseline=4mm,background rectangle/.style=
{fill=white},
   show background rectangle]
        \draw[shorten <>=2mm] (0,-1) -- (0,2);
        \draw[shorten <>=2mm] (-1,0) -- (1,0) -- (1,1) -- (-1,1);
        \node[tensor, minimum height=1.5cm] at (0,0.5) {$N$};
        \node[tensor] at (1,0.5) {$v_R$};
        \node at (2.0,0.5) {$=0$};
    \end{tikzpicture}  \quad \text{and} \qquad 
        \begin{tikzpicture}[baseline=4mm,background rectangle/.style=
{fill=white},
   show background rectangle]
        \draw[shorten <>=2mm] (0,-1) -- (0,2);
        \draw[shorten <>=2mm] (1,1) -- (-1,1) -- (-1,0) -- (1,0);
        \node[tensor, minimum height=1.5cm] at (0,0.5) {$N$};
        \node[tensor] at (-1,0.5) {$v_L$};
        \node at (1.5,0.5) {$=0$};
    \end{tikzpicture}  \ .
\end{align}
Then the MPO is a product operator (i.e.,  the virtual dimension of $A$ is one).\end{result}\end{shaded}
\subsection{The extended $R$-matrix and modified RLL relation}
If we transform a Yang--Baxter-integrable lattice model by conjugating by an MPO that has an exact MPO inverse,  we can reason as in \cref{sec:cluster} to arrive at the modified RLL relation. 

Indeed, starting with a Yang--Baxter-integrable model means that we have an $R$-matrix satisfying the standard RLL relation, and that we construct the transfer matrices from the corresponding Lax operators, $L(u)$. Let us consider an MPO with MPO inverse as defined around \cref{eq:ABMPO}. This MPO can be used to define a dual model where we have transformed Lax operators
\begin{align}
    \includegraphics[]{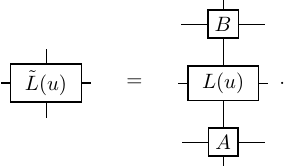} 
\end{align}
From the eigenvectors of the $A$-$B$ transfer matrix, 
we can then define an extended $R$-matrix as above by
\begin{align}
    \includegraphics[]{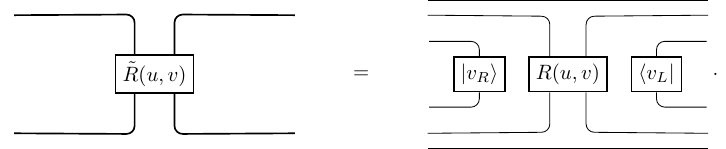} 
\end{align}
We block physical sites so that the nilpotent tensor has $n_0=1$. Then this $\tilde{R}$ and $\tilde{L}$ satisfy the modified RLL relation, as well as the identities \cref{eq:RLLP,eq:LLPLL} replacing $\ket{v_0}\bra{v_0}\rightarrow \ket{v_R}\bra{v_L}$ in $P$. This moreover gives us the modified YBA \cref{eq:modifiedYBA}. 

To see this, we can use \eqref{eq:mpo_transfer}
in an analogue of the analysis in \cref{app:RLL}. The nilpotent tensor acts as an obstruction and appears in the right-hand-side of the usual RLL relation as follows:
\begin{align}
\includegraphics[]{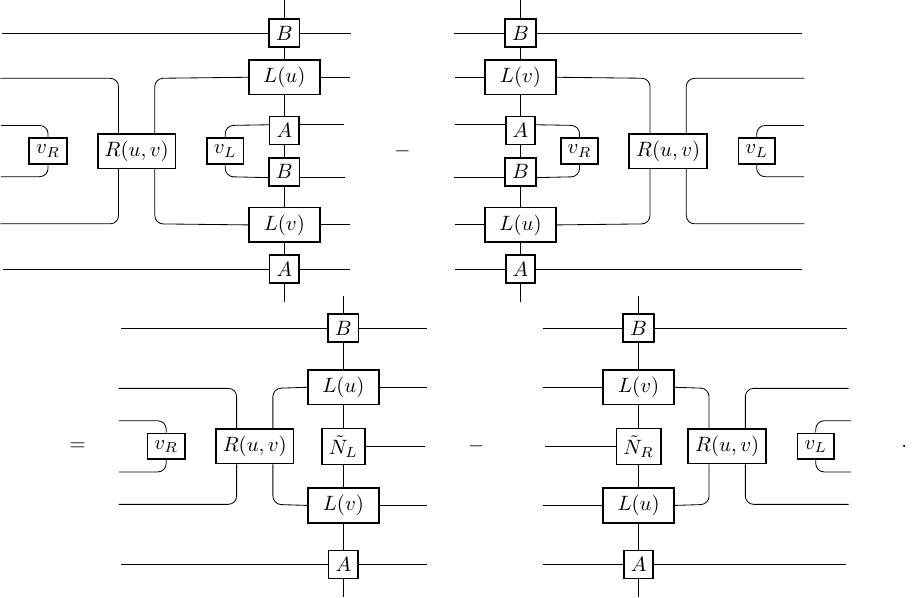} \label{eq:RLLgeneralised}
\end{align}
Here the tensors $\tilde{N}_L$ and $\tilde{N}_R$ are defined by 
\begin{align}\label{eq:mpu_simple_assumption1}
    \begin{tikzpicture}[baseline=4mm]
     \draw[shorten <>=2mm] (-2.5,-1) -- (-2.5,2);
          \draw[thick,shorten <>=2mm] (-2.5,0.5) -- (-3.5,0.5);
     \node[tensor, minimum height=1cm] at (-2.5,0.5) {$\tilde{N}_R$};
     \node (node3B) at (-1.5,0.5)  {$\equiv$};
        \draw[shorten <>=2mm] (0,-1) -- (0,2);
        \draw[shorten <>=2mm] (-1,0) -- (1,0) -- (1,1) -- (-1,1);
        \node[tensor, minimum height=1.5cm] at (0,0.5) {$N$};
        \node[tensor] at (1,0.5) {$v_R$};
    \end{tikzpicture} \quad \text{and} \quad 
    \begin{tikzpicture}[baseline=4mm]
     \draw[shorten <>=2mm] (-2.5,-1) -- (-2.5,2);
          \draw[thick,shorten <>=2mm] (-2.5,0.5) -- (-1.5,0.5);
     \node[tensor, minimum height=1cm] at (-2.5,0.5) {$\tilde{N}_L$};
     \end{tikzpicture}
     \equiv \quad
        \begin{tikzpicture}[baseline=4mm]
        \draw[shorten <>=2mm] (0,-1) -- (0,2);
        \draw[shorten <>=2mm] (1,1) -- (-1,1) -- (-1,0) -- (1,0);
        \node[tensor, minimum height=1.5cm] at (0,0.5) {$N$};
        \node[tensor] at (-1,0.5) {$v_L$};
    \end{tikzpicture}  \ .
\end{align}
Note that, as discussed in \cite{Jones22} for the MPU case, $\tilde{R}$ satisfies the YBE. It is not, however, a regular solution. In particular, it is not invertible. This non-invertibility causes no issue in the argument for commuting transfer matrices -- the proof is the same as in \cref{sec:cluster}.

\subsection{Obstructions and charge pumps}
Suppose that the right-hand-side of \cref{eq:RLLgeneralised} vanishes, meaning that our extended $R$-matrix satisfies a standard RLL relation. The simplest way this can happen is when $\tilde{N}_R=\tilde{N}_L=0$. From \cref{sec:MPUtrivial}, this implies that the MPO must be bond dimension one, so this vanishing cannot occur for non-trivial bond dimension. As discussed in \cref{subsec:invertibleonsite}, on-site transformations trivially preserve the integrable structure. 

This observation does not rule out more subtle ways that the usual RLL equation could be obtained for the extended $R$-matrix. For example, the contraction of the Lax operators with $\tilde{N}_{L/R}$ could vanish. We verify that this does not occur generically in the XXZ model case in \cref{app:RLL}. 

Another interesting point is the connection of the obstructions $\tilde{N}_{L/R}$ to charge pumps. 
From \cref{sec:cluster}, the decomposition \cref{eq:MPUdecomposition} makes clear that $\tr(M_1M_2\dots M_k) =\mathbb{I}$ for any $k$ (of course, this can be interpreted as the MPU generating a unitary for any system size). The trace is essential, since \begin{align}
   M_1\dots M_k = \ket{v_0}\bra{v_0} \mathbb{I}+ \ket{v_0}\bra{v_2} Z_k+
     \ket{v_1}\bra{v_0} Z_1+ \ket{v_1}\bra{v_2} Z_1Z_k \ . \label{eq:chargepump}
\end{align} 
Without the trace, this operator is a variant of ${U}_c$ that depends on the boundary termination on the virtual space. Let ${U}^{\textrm{OBC}}_c$ be ${U}_c$ on an open chain, where we remove the operator $CZ_{L,1}$ that connects the two ends. Fixing $\bra{v_0}M_1M_2\dots M_k \ket{v_0} $ amounts to applying ${U}_c^{\textrm{OBC}} ({U}^{\textrm{OBC}}_c)^{-1}=\mathbb{I}$.
On the other hand $\bra{v_1}M_1M_2\dots M_k \ket{v_2} $ corresponds to applying $\left({U}_c^{\textrm{OBC}}\right)^2 =Z_1 Z_L$. This can be interpreted as a charge pump for the $\prod_j X_j$ symmetry that is preserved when we map $\mathbf{H}_0$ to $\mathbf{H}_{\mathrm{cluster}}$ \cite{Tantivasadakarn23,Jones25b}. The charged operator, $Z$, obstructs our local RLL relation \cref{eq:RLL}, but must appear so that we have a possible boundary termination of \cref{eq:chargepump} such that we recover the pump behaviour of $\left({U}_c^{\textrm{OBC}}\right)^2$.

Given our motivation of using MPU transformations to change symmetry properties of our integrable model, we expect this behaviour to be typical. Indeed, following the discussion in \cite{Jones25b}, we expect that SPT entanglers will often give rise to charge pumps, and thus in this important class of examples the nilpotent operator $N$ will have some symmetry charge. It would be interesting to explore this further, and to understand how this charge may constrain the modified RLL relation.

\subsection{The non-invertible case}
Here we make some general remarks on the non-invertible duality over integrable models. 
The construction of \cref{sec:KW} can be generalized to other non-invertible duality transformations that correspond to the gauging of certain discrete symmetries of the Hamiltonian (and the $R$-matrix), where the integrability is still preserved in the dual model, while some global symmetries might differ~\cite{Bhardwaj:2017xup, Cao:2025qnc}. The dual model possesses symmetries described by $\mathrm{Rep}(G)$, which can be non-invertible when $G$ is non-Abelian.
Alternatively, integrable models with non-invertible categorical symmetry can be built using the method outlined in \cite{Aasen20, Eck24, Lootens_2023, Lootens_2024}. It would be interesting to compare the results using both approaches. The baxterization process indicates that the integrability structure remains the same in the bulk after the duality transformation. Meanwhile, the boundary conditions that are compatible with the non-invertible duality are given by the categorical data of the gauged discrete group $G$, which \emph{a priori} do not necessarily preserve the integrability. We do not elaborate on this issue in the current paper, analysis is deferred to future work.

Another useful observation is to consider the non-invertible duality transformation between two integrable models in their field theory limit. This can be understood as orbifolding \cite{Ginsparg88}. For example, the XXZ spin chain Hamiltonian \eqref{eq:XXZHamiltonian} in the regime $|\Delta| \leq 1$ flows to the $c=1$ free compactified boson CFT with radius $r=\sqrt{\frac{\pi}{2(\pi-\arccos \Delta)}}$~\cite{Alcaraz_PRL_1987, Ginsparg88}. The dual model \eqref{eq:IzzHamiltonian} thus flows to the same $c=1$ CFT with $r^\prime = \frac{r}{2}$ after the $U(1)/\mathbb{Z}_2$ orbifolding\footnote{Note that this orbifolding procedure does \emph{not} correspond to the orbifold branch of the $c=1$ moduli space~\cite{Ginsparg88}. If we instead gauge the $\mathbb{Z}_2^x$ symmetry of the XXZ spin chain in the manner described in \cref{sec:KW}, we shall arrive at the orbifold branch with the same radius in the CFT limit. See also the discussion in \cite{Thorngren21}.}. A similar check can be performed by analysing the partition functions of the corresponding vertex models after inserting the non-invertible duality defects~\cite{Eck24}.

\section{Outlook}\label{sec:outlook}
In this work we analysed MPO dualities between integrable lattice models. For MPO with exact MPO inverse, we found that the MPO-dual $\check{R}$ matrix continues to satisfy the circuit YBE and therefore may be used in the baxterization approach. Care must be taken with boundary conditions in this case. For Yang--Baxter integrable models, we found that generically neither the $R$-matrix nor the local RLL equation is preserved. However, we could find a canonical extended $R$-matrix (using a projector derived from the MPO) that gives rise to a modified RLL equation and Yang--Baxter algebra. This allows for a local derivation of commuting transfer matrices in the dual model, and elucidates the integrable structure. In the case of non-invertible MPOs we found that the integrable structure of the dual model remains the same by virtue of the baxterization, and is reminiscent of the renowned vertex-face correspondence. We wrote down explicitly the transfer matrix of the dual model as a split-index MPO for the example of applying the KW duality to the XXZ spin chain. 

For MPU dualities the analysis ended up being identical to the more general case of an MPO with exact MPO inverse. It would be interesting to identify special classes of unitaries where the dualities are simplified, but not so simple as the on-site case. While we excluded one route to simplification above, it would be worth exploring the case of dual unitary transformations \cite{Piroli20}.

For non-invertible MPOs we focused on the case of discrete gauging. It would be very interesting to understand broader classes of models which could give useful, physically motivated, dualities of integrable models. It would be useful to understand more deeply the connection between classification of medium-range integrable models \cite{Gombor21} and those that arise by MPO transformation of nearest-neighbour models. Meanwhile, integrability is typically associated with an underlying quantum group structure. Although we have shown that baxterization is compatible with the duality transformations, the precise manner in which the quantum group structure is transformed under these dualities remains elusive and is left for future study.

More broadly, it would be very interesting to utilise modern tensor network techniques to gain a better understanding of the field of integrability. Exciting recent work in this direction \cite{Fendley25} that showed the XYZ spin chain is integrable via a tensor network approach, avoiding complications of the standard method, is encouraging.
Results for MPO with exact MPO inverse that we include above, such as locality of the logarithmic derivatives, may be useful in numerical approaches.

\section*{Acknowledgements}
We are grateful to Paul Fendley and Noah Linden for helpful discussions. The work of Y.M. is supported by the World Premier International Research Center Initiative (WPI), MEXT, Japan and the UTokyo Global Activity Support Program for Young Researchers. Y.M. is grateful for the hospitality of the Rudolf Peierls Centre for Theoretical Physics and St. John's College at the University of Oxford. A. M.
acknowledges funding by the Austrian Science Fund FWF (Grant
Nos.~\href{https://doi.org/10.55776/COE1}{10.55776/COE1},
\href{https://doi.org/10.55776/F71}{10.55776/F71},
\href{https://doi.org/10.55776/P36305}{10.55776/P36305}), by the European Union
-- NextGenerationEU, and by the European Union’s Horizon 2020 research and
innovation programme through Grant No.~863476 (ERC-CoG \mbox{SEQUAM}).

\appendix
\section{Proofs of matrix-product operator results}\label{app:MPO}
In this appendix we prove the results of \cref{sec:MPO}, taking the setup and statements in that section as a given.
\subsection{$A$-$B$ transfer matrix decomposition}
By assumption, $O_n(A) \cdot O_n(B) = \mathbb{I}^{\otimes n}$, or graphically, 
\begin{align*}
    O_n(A) O_n (B) = \ 
    \begin{tikzpicture}[baseline = 2.5mm, yscale=0.7]
        \draw (-0.9,1) rectangle (3.9, -1.4);
        \draw (-0.7,0) rectangle (3.7, -1.2);
        \foreach \x in {0,1,3} {
            \draw (\x,-1) --++ (0,3);
            \node[tensor] at (\x,1) {$A$};
            \node[tensor] at (\x,0) {$B$};
        }
        \node[fill=white] at (2,1) {$\dots$};
        \node[fill=white] at (2,0) {$\dots$};
    \end{tikzpicture} \ = \mathbb{I}^{\otimes n}.
\end{align*}
Both left and right hand side of this equation is a matrix product state (operator), and the right hand side ($\mathbf{I}^{\otimes n}$) is an \emph{injective} MPS (as it is a tensor product). Thus we can apply Proposition 20 and 21 from Ref.~\cite{Molnar18}. We directly obtain the results contained in \cref{eq:mpo_transfer,eq:n-orthogonal,eq:nilpotency}. 
\subsection{Locality preservation}
To prove the statement, notice first that outside of the support of $X$ the tensor $A$ is contracted with the tensor $B$ and results in the $A$-$B$ transfer matrix:
\begin{align}
    O_n(A) \cdot X \cdot O_n(B) = 
    \begin{tikzpicture}[baseline = 4mm]
        \draw (-0.9,2) rectangle (8.9, -1.4);
        \draw (-0.7,0) rectangle (8.7, -1.2);
        \foreach \x in {0,2,3, 5, 6, 8} {
            \draw (\x,-0.7) -- (\x,2.7);
            \node[tensor] at (\x,0) {$B$};
            \node[tensor] at (\x,2) {$A$};
        }
        \foreach \x in {1,4,7}{
            \node[fill=white] at (\x,2) {$\dots$};
            \node[fill=white] at (\x,0) {$\dots$};
        }
        \node[tensor, minimum width = 2.5cm] at (4,1) {$X$}; 
    \end{tikzpicture}     \ .
\end{align}
Outside of the support of $X$ we can thus use \cref{eq:mpo_transfer} and obtain, if $n>k+2n_0$, that
\begin{align}
    O_n(A) \cdot X \cdot O_n(B) = \sum_{l,m=0}^{n_0} \mathbb{I}^{\otimes (l_0-l)} \otimes 
    \begin{tikzpicture}[baseline = 9mm]
        \draw (-0.7,2) rectangle (8.7, 0);
        \foreach \x in {3, 5} {
            \draw (\x,-0.7) -- (\x,2.7);
            \node[tensor] at (\x,0) {$B$};
            \node[tensor] at (\x,2) {$A$};
        }
        \foreach \x in {0,2, 6, 8} {
            \draw (\x,-0.7) -- (\x,2.7);
            \node[tensor, minimum height=2.5cm] at (\x,1) {$N$};
        }
        \foreach \x in {1,4,7}{
            \node[fill=white] at (\x,2) {$\dots$};
            \node[fill=white] at (\x,0) {$\dots$};
        }
        \node[tensor, minimum width = 2.5cm] at (4,1) {$X$};
        \node[tensor] at (-0.7,1) {$R$};
        \node[tensor] at (8.7,1) {$L$};
    \end{tikzpicture}  \otimes \mathbb{I}^{\otimes r_0-r},   
\end{align}
where we have used that the single term containing only the tensor $N$ disappears, as it contains the contraction of $n-k>n_0$ tensors, which is zero. To draw the diagram, we also assumed that $n=l_0+k+r_0$, the operator $X$ is located at sites $(l_0+1), \dots, (l_0+k)$, and that $l_0>n$ and $r_0>n$ (that is, $n$ is large enough and that $X$ is somewhere in the `middle'). The resulting operator is clearly local: all the terms in the sum are localized between the sites $l_0-n_0$ and $l_0+k+n_0$. 

\subsection{Generator of local charges} 
Similar to the argument above, we can show that the logarithmic derivative of such an MPO is local. 

For that, note first that the derivative of $O_n(A)$ can be written as 
\begin{align}
    \partial_t O_n(A_t) = \sum_i 
    \begin{tikzpicture}[baseline = -1mm, yscale=0.6]
        \draw (-0.7,0) rectangle (5.7, -1.2);
        \foreach \x in {0,1,5} {
            \draw (\x,-1) --++ (0,2);
            \node[tensor] at (\x,0) {$A$};
        }
        \draw (3,-1) --++ (0,2);
        \node[tensor] at (3,0) {$X$};
        \node[fill=white] at (2,0) {$\dots$};
        \node[fill=white] at (4,0) {$\dots$};
    \end{tikzpicture} \ ,
\end{align}
where the tensor $X = \partial_t A_t$ appears on the $i^{th}$ site. 

Using this formula, the logarithmic derivative can be written as
\begin{align}
    \partial_t O_n(A_t) \cdot B_t = \sum_i
    \begin{tikzpicture}[baseline = -1mm, yscale=0.8]
        \draw (-0.9,1) rectangle (5.9, -1.4);
        \draw (-0.7,0) rectangle (5.7, -1.2);
        \foreach \x in {0,1,5} {
            \draw (\x,-1) --++ (0,3);
            \node[tensor] at (\x,1) {$A$};
            \node[tensor] at (\x,0) {$B$};
        }
        \draw (3,-1) --++ (0,3);
        \node[tensor] at (3,1) {$A$};
        \node[tensor] at (3,0) {$X$};
        \node[fill=white] at (2,0) {$\dots$};
        \node[fill=white] at (4,0) {$\dots$};
        \node[fill=white] at (2,1) {$\dots$};
        \node[fill=white] at (4,1) {$\dots$};
    \end{tikzpicture} \ .    
\end{align}
Finally, by repeating the argument in the previous section, each term in the sum is local.

\subsection{Triviality when $N$ annihilates the dominant left-right eigenvectors}
Notice first that for any $n$, the product of $n$ consecutive MPO tensors together with the inverse MPO tensors can be written as
\begin{align}
    \begin{tikzpicture}[baseline=4mm]
        \draw[shorten <>=2mm] (-1,0) -- (3,0);
        \draw[shorten <>=2mm] (-1,1) -- (3,1);
        \draw[shorten <>=2mm] (0,-1) -- (0,2);
        \draw[shorten <>=2mm] (2,-1) -- (2,2);
        \node[tensor] at (0,0) {$B$};
        \node[tensor] at (2,0) {$B$};
        \node[tensor] at (0,1) {$A$};
        \node[tensor] at (2,1) {$A$};
        \node[fill=white] at (1,0) {$\dots$};
        \node[fill=white] at (1,1) {$\dots$};
    \end{tikzpicture} = 
    \begin{tikzpicture}[baseline=4mm, xscale=0.7]
        \draw[shorten <>=2mm] (0,-1) -- (0,2);
        \draw[shorten <>=2mm] (2,-1) -- (2,2);
        \node[fill=white] at (1,0.5) {$\dots$};
        \draw[shorten <>=2mm] (-2,1) -- (-1,1) -- (-1,0) -- (-2,0);
        \draw[shorten <>=2mm] (4,1) -- (3,1) -- (3,0) -- (4,0);
        \node[tensor] at (-1,0.5) {$L$};
        \node[tensor] at (3,0.5) {$R$};
    \end{tikzpicture} +
    \begin{tikzpicture}[baseline=4mm]
        \draw[shorten <>=2mm] (0,-1) -- (0,2);
        \draw[shorten <>=2mm] (2,-1) -- (2,2);
        \draw[shorten <>=2mm] (-1,0) -- (3,0);
        \draw[shorten <>=2mm] (-1,1) -- (3,1);
        \node[tensor, minimum height=1.5cm] at (0,0.5) {$N$};
        \node[tensor, minimum height=1.5cm] at (2,0.5) {$N$};
        \node[fill=white] at (1,0) {$\dots$};
        \node[fill=white] at (1,1) {$\dots$};
    \end{tikzpicture} ,
\end{align}
where we have expanded the left-hand-side using \cref{eq:mpo_transfer} and noticed that using \cref{eq:mpu_simple_assumption} all but two terms cancel in the resulting sum.

Since $N$ is nilpotent, for sufficiently large $n$, we have 
\begin{align}
    \begin{tikzpicture}[baseline=4mm]
        \draw[shorten <>=2mm] (-1,0) -- (3,0);
        \draw[shorten <>=2mm] (-1,1) -- (3,1);
        \draw[shorten <>=2mm] (0,-1) -- (0,2);
        \draw[shorten <>=2mm] (2,-1) -- (2,2);
        \node[tensor] at (0,0) {$B$};
        \node[tensor] at (2,0) {$B$};
        \node[tensor] at (0,1) {$A$};
        \node[tensor] at (2,1) {$A$};
        \node[fill=white] at (1,0) {$\dots$};
        \node[fill=white] at (1,1) {$\dots$};
    \end{tikzpicture} = 
    \begin{tikzpicture}[baseline=4mm, xscale=0.7]
        \draw[shorten <>=2mm] (0,-1) -- (0,2);
        \draw[shorten <>=2mm] (2,-1) -- (2,2);
        \node[fill=white] at (1,0.5) {$\dots$};
        \draw[shorten <>=2mm] (-2,1) -- (-1,1) -- (-1,0) -- (-2,0);
        \draw[shorten <>=2mm] (4,1) -- (3,1) -- (3,0) -- (4,0);
        \node[tensor] at (-1,0.5) {$L$};
        \node[tensor] at (3,0.5) {$R$};
    \end{tikzpicture} .
\end{align}

Let us consider this equation with the boundary $X \otimes \mathbb{I}$, where $\mathbb{I}$ acts on the lower virtual index, and $X$ is an arbitrary matrix acting on the upper virtual index. We obtain
\begin{align}
    O_n(A,X) \cdot O_n(B) = \tr(LRX) \cdot \mathbb{I}^{\otimes n}\ ,
\end{align}
where 
\begin{align}
    O_n(A,X) = \sum_{ij} \tr (X A_{i_1,j_1} A_{i_2,j_2} \dots A_{i_n,j_n}) \ket{i_1 i_2  \dots i_n} \bra{j_1 j_2 \dots j_n}\ ,
\end{align}
and thus $O_n(A,X) \propto O_n(A)$ for all $X$. As $A$ is an injective tensor, this implies that its bond dimension is one, and thus $O_n(A) = o^{\otimes n}$ for a local invertible operator $o$.

\section{The transformed RLL relation}\label{app:RLL}
In this appendix we analyse the local RLL relation for an MPU transformed model after including the projector onto $\ket{v_0}$ in the virtual space. We use the particular expansion given in \cref{eq:MPUdecomposition} for the cluster entangler, but this is representative of the general case. 
\begin{align}
\includegraphics[]{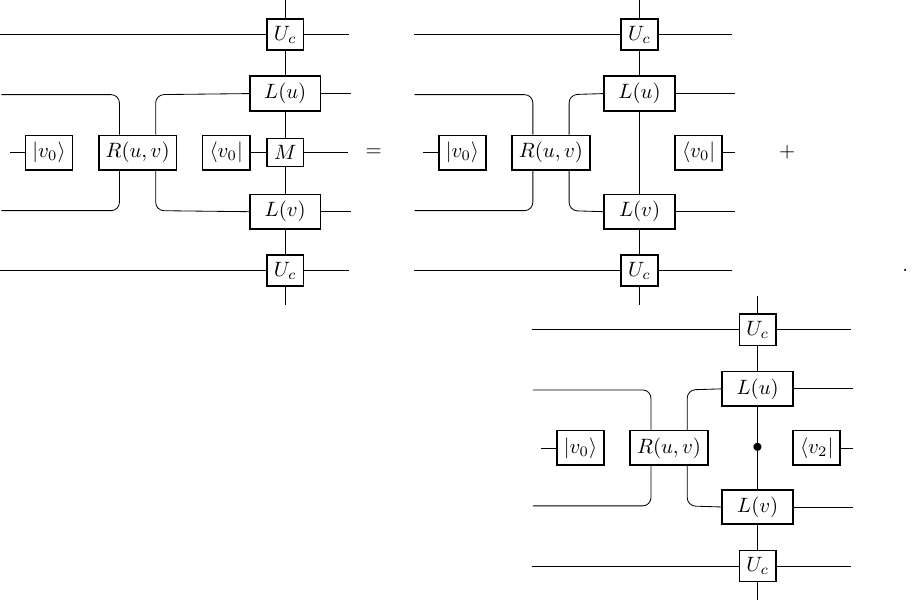}
\end{align}
In the first term on the right-hand-side, we can now use the standard RLL relation \eqref{eq:RLLgeneral}. The second term contains a $Z$ operator, and there is no reason that $R(u,v)$ acts as an intertwiner in such a case (and for the XXZ model it does not, see below for the explicit calculation). We thus have that
\begin{align}
\includegraphics[]{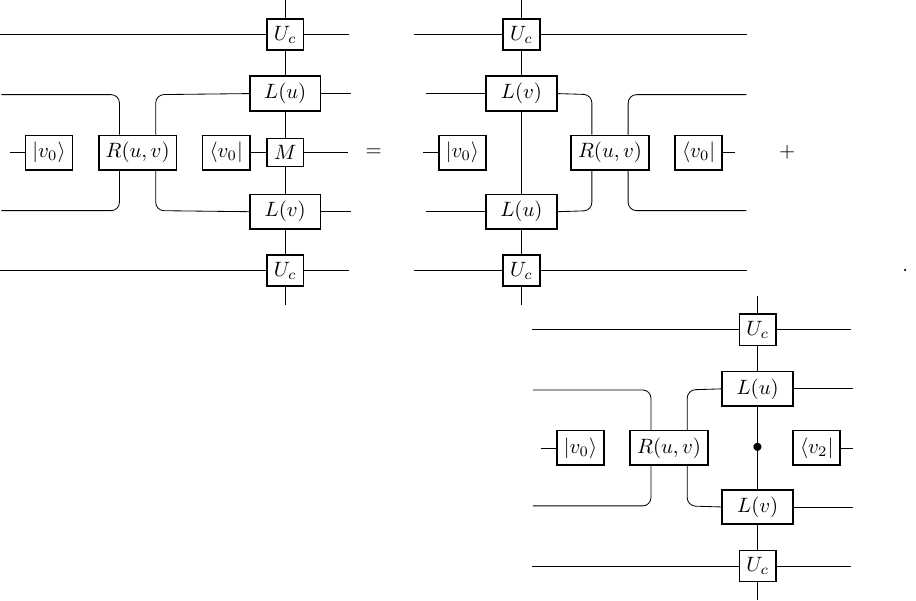} 
\end{align}
Analysing the right-hand-side of the RLL relation, we have that
\begin{align}
\includegraphics[]{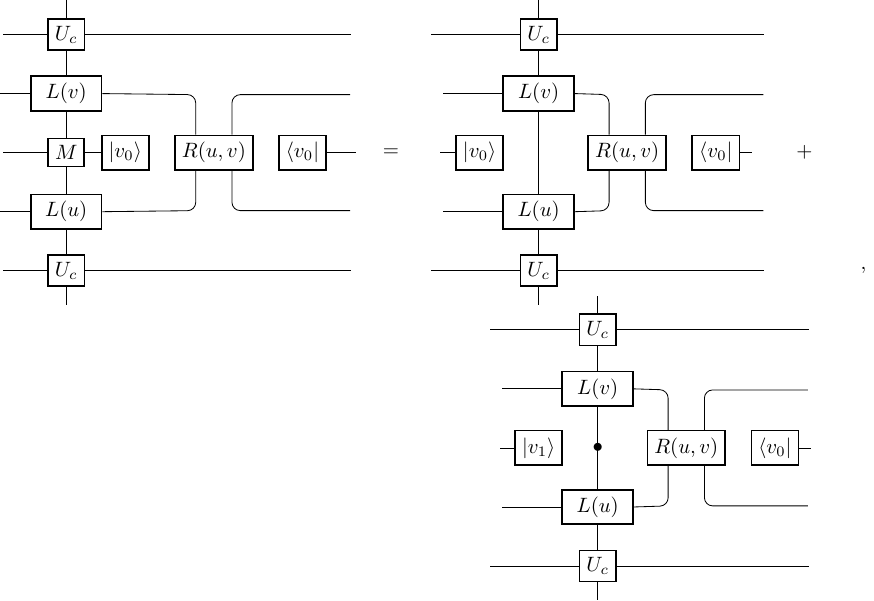} 
\end{align}
taking the difference gives \cref{eq:RLL}. 

Since the vectors $|v_j \rangle$ are mutually orthogonal, the right hand-side of \cref{eq:RLL} is independent of the left-hand side. Moreover, we expect that each side of \cref{eq:RLL} is non-vanishing for any $u,v \in \mathbb{C}$. In the XXZ case, we have that

\begin{align}
 \includegraphics[]{Eq103}& \\ \nonumber
& = \begin{pmatrix}
    M_{1} & {\color{gray} \cdot } & {\color{gray} \cdot } & {\color{gray} \cdot } & {\color{gray} \cdot} & {\color{gray} \cdot} & {\color{gray} \cdot} & {\color{gray} \cdot} \\
    {\color{gray} \cdot} & M_2 & M_3 &  {\color{gray} \cdot} & M_4 & {\color{gray} \cdot} & {\color{gray} \cdot} & {\color{gray} \cdot} \\
    {\color{gray} \cdot} & M_5 & M_6 &  {\color{gray} \cdot} & M_7 & {\color{gray} \cdot} & {\color{gray} \cdot} & {\color{gray} \cdot}  \\
    {\color{gray} \cdot} & {\color{gray} \cdot} & {\color{gray} \cdot} & M_8 & {\color{gray} \cdot} & M_{9} & M_{10} & {\color{gray} \cdot}  \\
    {\color{gray} \cdot} & M_{11} & M_{12} &  {\color{gray} \cdot} & M_{13} & {\color{gray} \cdot} & {\color{gray} \cdot} & {\color{gray} \cdot}  \\
    {\color{gray} \cdot} & {\color{gray} \cdot} & {\color{gray} \cdot} & M_{14} & {\color{gray} \cdot} & M_{15} & M_{16} & {\color{gray} \cdot}  \\
    {\color{gray} \cdot} & {\color{gray} \cdot} & {\color{gray} \cdot} & M_{17} & {\color{gray} \cdot} & M_{18} & M_{19} & {\color{gray} \cdot}  \\
    {\color{gray} \cdot} & {\color{gray} \cdot} & {\color{gray} \cdot} & {\color{gray} \cdot} & {\color{gray} \cdot} & {\color{gray} \cdot} & {\color{gray} \cdot} & -M_1
\end{pmatrix} \; ,
\end{align}
where the matrix elements are given explicitly as
\begin{align}
\begin{split}
    & M_1 = \sin (u-v+\mu) \sin (u+\mu) \sin (v+\mu) \; , \\
    & M_2 = -\sin u \, \sin v \sin (u-v+\mu) \; , \quad M_3 =   -\sin u \, \sin \mu \sin (u-v+\mu) \; , \\
    & M_4 = \sin \mu \, \sin (v+\mu) \sin (u-v+\mu) \; ,\\
    & M_5 = \sin \mu \, (\sin (u+ \mu ) \sin (u-v)-\sin \mu \, \sin v ) \; , \\
    & M_6 = \sin v \, \sin (u+\mu ) \sin (u-v) - \sin ^3 \mu \; , \\
    & M_7 = \sin \mu \, \sin u \, \sin (v+\mu ) \; , \quad M_8 = -\sin u \, \sin (u-v) \sin (v+\mu) \; , \\
    & M_9 = \sin \mu (\sin \mu \, \sin (u-v)-\sin v \, \sin (u+\mu)) \; , \\
    & M_{10} = \sin \mu (\sin v \, \sin (u-v) - \sin \mu \, \sin (u+\mu) ) \; , \\
    & M_{11} = \sin \mu (\sin \mu \, \sin (u+ \mu) -\sin v \, \sin (u-v)) \; , \\
    & M_{12} = \sin \mu (\sin v \, \sin (u+ \mu) - \sin \mu \sin (u-v) ) \; , \\ 
    & M_{13} = \sin u \, \sin (u-v) \sin (v+\mu ) \; , \quad M_{14} = -\sin u \, \sin \mu \,\sin (v+\mu ) \; , \\
    & M_{15} = \sin^3 \mu - \sin v \sin (u+\mu ) \sin (u-v) \; , \\
    & M_{16} = \sin \mu (\sin \mu \, \sin v - \sin (u+\mu ) \sin (u-v)) \; , \\
    & M_{17} = -\sin \mu \, \sin (v+\mu ) \sin (u-v+\mu) \; , \\
    & M_{18} = \sin \mu \, \sin u \,  \sin (u-v + \mu ) \; , \quad M_{19} = \sin u \, \sin v \,  \sin (u-v + \mu ) \; .
\end{split}
\end{align}
(Recall that each index in the diagram corresponds to a two-dimensional space. For the matrix representation the row corresponds to grouping anticlockwise the left virtual indices and the bottom physical index, while the column corresponds to grouping anticlockwise the remaining indices.)

Meanwhile, the term
\begin{align}
\includegraphics[]{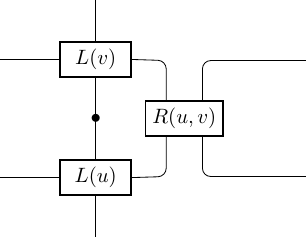} &\nonumber \\
& = \begin{pmatrix}
    M_{1} & {\color{gray} \cdot } & {\color{gray} \cdot } & {\color{gray} \cdot } & {\color{gray} \cdot} & {\color{gray} \cdot} & {\color{gray} \cdot} & {\color{gray} \cdot} \\
    {\color{gray} \cdot} & M_2 & M_3 &  {\color{gray} \cdot} & M_4 & {\color{gray} \cdot} & {\color{gray} \cdot} & {\color{gray} \cdot} \\
    {\color{gray} \cdot} & M_5 & M_6 &  {\color{gray} \cdot} & M_7 & {\color{gray} \cdot} & {\color{gray} \cdot} & {\color{gray} \cdot}  \\
    {\color{gray} \cdot} & {\color{gray} \cdot} & {\color{gray} \cdot} & M_8 & {\color{gray} \cdot} & M_{9} & M_{10} & {\color{gray} \cdot}  \\
    {\color{gray} \cdot} & M_{11} & M_{12} &  {\color{gray} \cdot} & M_{13} & {\color{gray} \cdot} & {\color{gray} \cdot} & {\color{gray} \cdot}  \\
    {\color{gray} \cdot} & {\color{gray} \cdot} & {\color{gray} \cdot} & M_{14} & {\color{gray} \cdot} & M_{15} & M_{16} & {\color{gray} \cdot}  \\
    {\color{gray} \cdot} & {\color{gray} \cdot} & {\color{gray} \cdot} & M_{17} & {\color{gray} \cdot} & M_{18} & M_{19} & {\color{gray} \cdot}  \\
    {\color{gray} \cdot} & {\color{gray} \cdot} & {\color{gray} \cdot} & {\color{gray} \cdot} & {\color{gray} \cdot} & {\color{gray} \cdot} & {\color{gray} \cdot} & -M_1
\end{pmatrix}^{\mathrm{T}} \; ,
\end{align}
which implies that for generic values of $u,v \in \mathbb{C}$, we have that
\begin{align}
\vcenter{\hbox{
    \includegraphics[]{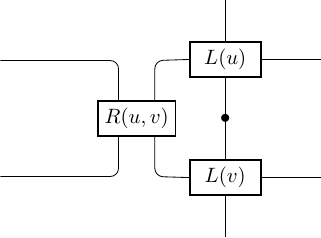}}}-\vcenter{\hbox{
    \includegraphics[]{Eq105.pdf}}}
\neq 0 \; .
\end{align}
Therefore, at least for the XXZ case, \cref{eq:RLL} is non-vanishing for generic values of $u,v \in \mathbb{C}$.
\bibliography{v2.bbl}{}

\end{document}